\documentclass[acmsmall,screen]{acmart}
\pdfoutput=1
\setcopyright{rightsretained}
\acmPrice{}
\acmDOI{10.1145/3485492}
\acmYear{2021}
\copyrightyear{2021}
\acmSubmissionID{oopsla21main-p81-p}
\acmJournal{PACMPL}
\acmVolume{5}
\acmNumber{OOPSLA}
\acmArticle{115}
\acmMonth{10}

\usepackage{booktabs}   
\usepackage{subcaption} 
\usepackage[utf8]{inputenc} 
\usepackage[T1]{fontenc}    
\usepackage{amsfonts}       
\usepackage{nicefrac}       
\usepackage{microtype}      
\usepackage{amsthm}
\usepackage{xspace}
\usepackage{cleveref}
\usepackage{multirow}
\usepackage{calc}

\usepackage{tikz}
\usetikzlibrary{positioning}
\usetikzlibrary{calc}

\usepackage{wrapfig}
\usepackage{listings}
\usepackage{mathpartir}
\usepackage{xargs}
\usepackage{pifont}
\usepackage{balance}
\usepackage{enumitem}
\lstdefinelanguage{muF}
{morekeywords={
     node,proba,rec,and,where,
     val,fun,pre,
     let,in,if,then,else,
     sample, observe, factor, infer, eval,
     stream, unfold, init, step, reset
   },
 sensitive=true,
 morecomment=[n]{(*}{*)},
 morestring=[b]",
 escapechar=\%,
 columns=fullflexible,
 keepspaces=true,
 basicstyle=\ttfamily,
 mathescape=true,
}
\newcommand{\dsapp}[2]{\ensuremath{\textrm{app}(#1, #2)}}

\newcommand{\dsassume}{\ensuremath{\mathit{assume}}}
\newcommand{\dsobserve}{\ensuremath{\mathit{observe}}}
\newcommand{\dsvalue}{\ensuremath{\mathit{value}}}
\newcommand{\dsdraw}{\ensuremath{\mathit{draw}}}
\newcommand{\dsdistribution}{\ensuremath{\mathit{distribution}}}
\newcommand{\dsinfer}{\ensuremath{\mathit{infer}}}

\newcommand{\dsllgraft}{\ensuremath{\mathit{graft}}}

\newcommand{\samplearrow}{\leftsquigarrow}

\newcommand*{\SavedLstInline}{}
\LetLtxMacro\SavedLstInline\lstinline
\DeclareRobustCommand*{\lstinline}{%
  \ifmmode
    \let\SavedBGroup\bgroup
    \def\bgroup{%
      \let\bgroup\SavedBGroup
      \hbox\bgroup
    }%
  \fi
  \SavedLstInline
}

\def\zl{\lstinline[basicstyle=\normalsize\ttfamily]}
\def\zlm{\lstinline[basicstyle=\small\ttfamily]}

\tikzstyle{initialized}=[circle, draw=black, minimum width=1ex]
\tikzstyle{marginalized}=[circle, draw=black, fill=gray!30, minimum width=0.2cm]
\tikzstyle{realized}=[circle, draw=black, fill=gray!80, minimum width=0.2cm]
\newcommandx{\link}[4][1={-stealth}, 4={}]{\draw[#1] (#2) to[bend left=25] node[pos=0.5] {#4} (#3)}
\newcommand{\xmark}{\text{\ding{55}}}
\newcommand{\rmark}{{\Large{\color{red!90!black}\text{\ding{55}}}}}

\newcommand{\Comment}[1]{}
\newcommand{\ProbZelus}[0]{ProbZelus\xspace}

\begin{document}

\title{Statically Bounded-Memory Delayed Sampling for Probabilistic Streams}         


\author{Eric Atkinson}
\affiliation{
  \institution{MIT}            
  \country{USA}                    
}

\author{Guillaume Baudart}
\affiliation{
  \institution{INRIA, École normale supérieure -- PSL University}           
  \country{France}                   
}

\author{Louis Mandel}
\affiliation{
  \institution{MIT-IBM Watson AI Lab, IBM Research}            
  \country{USA}                    
}
\author{Charles Yuan}
\affiliation{
  \institution{MIT}            
  \country{USA}                    
}
\author{Michael Carbin}
\affiliation{
  \institution{MIT}            
  \country{USA}                    
}

\newcommand{\dsnode}{{X}}
\newcommand{\progvar}{v}
\newcommand{\dsgraph}{g}
\newcommand{\DSGraph}{G}
\newcommand{\RVSet}{\mathcal{X}}
\newcommand{\RVUniv}{RV}

\newcommand{\freerandvars}{\mathit{frv}}
\newcommand{\fresh}{\mathit{fresh}}
\newcommand{\conjugate}{\mathit{conj}}
\newcommand{\reachable}{\mathit{reachable}}

\newcommand{\trassume}{\overline{\dsassume}}
\newcommand{\trvalue}{\overline{\dsvalue}}
\newcommand{\trobserve}{\overline{\dsobserve}}

\newcommand{\trace}{\tau}
\newcommand{\traceeval}{\ensuremath{\texttt{eval}}}
\newcommand{\traceobs}{\ensuremath{\texttt{obs}}}

\newcommand{\type}{t}

\newcommand{\muF}{\ensuremath{\mu F}\xspace}

\newcommand{\op}{\ensuremath{\mathit{op}}}

\newcommand{\muFval}[2]{\ensuremath{\texttt{val}\ #1\ \texttt{=}\ #2}}
\newcommand{\muFfun}[3]{\ensuremath{\texttt{val}\ #1\ \texttt{=}\ \texttt{fun }#2\texttt{ -> }#3}}

\newcommand{\muFpair}[2]{\ensuremath{\texttt{(}#1\texttt{,}#2\texttt{)}}}
\newcommand{\muFop}[1]{\ensuremath{\op\texttt{(}#1\texttt{)}}}
\newcommand{\muFapp}[2]{\ensuremath{#1\texttt{(}#2\texttt{)}}}
\newcommand{\muFif}[3]{\ensuremath{\texttt{if}\ #1\ \texttt{then}\ #2\ \texttt{else}\ #3}}
\newcommand{\muFletin}[3]{\ensuremath{\texttt{let }#1\texttt{ = }#2\texttt{ in }#3}}
\newcommand{\muFsample}[1]{\ensuremath{\texttt{sample(}#1\texttt{)}}}
\newcommand{\muFobserve}[2]{\ensuremath{\texttt{observe(}#1\texttt{,}#2\texttt{)}}}
\newcommand{\muFinfer}[1]{\ensuremath{\texttt{infer}\texttt{(}{#1}\texttt{)}}}
\newcommand{\muFreset}[1]{\ensuremath{\texttt{reset}\texttt{(}{#1}\texttt{)}}}
\newcommand{\muFstream}[4]{\ensuremath{\texttt{stream}\ \texttt{\{}\ \texttt{init}\ \texttt{=}\ #1\ \texttt{;}\ \texttt{step}\muFpair{#2}{#3}\ \texttt{=}\ #4\ \texttt{\}}}}
\newcommand{\muFinit}[1]{\ensuremath{\texttt{init}\texttt{(}{#1}\texttt{)}}}
\newcommand{\muFunfold}[2]{\ensuremath{\texttt{unfold}\muFpair{#1}{#2}}}

\newcommand{\typeD}{\ensuremath{\texttt{D}}\xspace}
\newcommand{\typeP}{\ensuremath{\texttt{P}}\xspace}
\newcommand{\typekind}[4]{\ensuremath{{#1} \vdash^{#4} {#2} : {#3}}}

\newcommand{\typebool}{\ensuremath{\texttt{bool}}}
\newcommand{\typefloat}{\ensuremath{\texttt{float}}}
\newcommand{\typeunit}{\ensuremath{\texttt{unit}}}
\newcommand{\typetimes}[2]{{#1} \times {#2}}
\newcommand{\typefun}[2]{{#1} \rightarrow {#2}}
\newcommand{\typenode}[4]{{#1} \Rightarrow_{#2}^{#3} {#4}}
\newcommand{\typeinstance}[4]{{#1} \Rrightarrow_{#2}^{#3} {#4}}

\newcommand{\typedist}[1]{\ensuremath{{#1}\ \texttt{dist}}}
\newcommand{\typesampler}[1]{\ensuremath{{#1}\ \texttt{dist}}}

\newcommand{\kindof}[1]{\ensuremath{\mathit{kindOf}(#1)}}
\newcommand{\typeof}[1]{\ensuremath{\mathit{typeOf}(#1)}}

\newcommand{\sem}[1]{\llbracket #1 \rrbracket}
\newcommand{\psem}[1]{\{\mkern-3.8mu[ #1 ]\mkern-3.8mu\}}
\newcommand{\indicator}{\mathbb{1}}
\newcommand{\letin}[1]{\ensuremath{\mathit{let} \; #1 \; \textit{in} \;}}
\newcommand{\fun}[1]{\ensuremath{\lambda #1. \;}}

\newcommand{\stinitialized}{\mathit{Initialized}}
\newcommand{\stmarginalized}{\mathit{Marginalized}}
\newcommand{\strealized}{\mathit{Realized}}

\newcommand{\coreinit}{\mathit{initialize}}
\newcommand{\coremarg}{\mathit{marginalize}}
\newcommand{\corecond}{\mathit{condition\_with\_value}}
\newcommand{\coredraw}{\mathit{draw}}
\newcommand{\corerealize}{\mathit{realize}}
\newcommand{\coreparent}{\mathit{get\_parent}}
\newcommand{\corechild}{\mathit{get\_child}}

\newcommand{\infrule}[2]{
  {\begin{array}[b]{cc}
      {#1} \vspace{-.2cm}\\
      \hrulefill\\
      {#2}
   \end{array}}}
\newcommand{\axiom}[1]{
    {\begin{array}[b]{cc}
        \hrulefill\\
        {#1}
     \end{array}}}

\begin{abstract}
{\em Probabilistic programming languages} aid developers performing Bayesian inference.
These languages provide programming constructs and tools for probabilistic modeling and automated inference.
Prior work introduced a probabilistic programming language, {\ProbZelus}, to extend probabilistic programming functionality to unbounded streams of data.
This work demonstrated that the {\em delayed sampling} inference algorithm could be extended to work in a streaming context.  
\ProbZelus showed that while delayed sampling could be effectively deployed on some programs, depending on the probabilistic model under consideration, delayed sampling is not guaranteed to use a bounded amount of memory over the course of the execution of the program.

In this paper, we present conditions on a probabilistic program's execution under which delayed sampling will execute in bounded memory.
The two conditions are dataflow properties of the core operations of delayed sampling: the {\em $m$-consumed property} and the {\em unseparated paths property}.
A program executes in bounded memory under delayed sampling if, and only if, it satisfies the $m$-consumed and unseparated paths properties.
We propose a static analysis that abstracts over these properties to soundly ensure that any program that passes the analysis satisfies these properties, and thus executes in bounded memory under delayed sampling.

\end{abstract}




\maketitle

\section{Introduction}

{\em Probabilistic programming languages} aid developers performing Bayesian inference
\cite{narayanan2016probabilistic,augurv2,church,edward,gen,pyro,r2,shuffle,stan,venture,webppl,tabular,figaro,turing,blog}.
These languages provide programming constructs and tools for probabilistic modeling and automated inference.
Researchers have developed probabilistic programming languages for several domains, including data science~\cite{stan}, machine learning~\cite{pyro,edward},
scientific simulation~\cite{etalumis}, and real-time control~\cite{probzelus}.

\paragraph{Probabilistic Programming with Streams}
In this paper, we consider programs that accept inputs and compute outputs at discrete time steps, with
the outputs of each step flowing into the environment to affect future inputs to the program. 
Mathematically, one can model these programs as computations that operate on and produce infinite {\em streams}.
Computing with streams is a common computational model for applications in real-time control, such as robotics and avionics~\cite{colaco-2017}.
For example, control for an airplane fly-by-wire system can be implemented as a program transforming a stream of altitude measurements into a stream of commands to the engine. 

\citet{probzelus} introduced a probabilistic programming language, {\ProbZelus}, to enable probabilistic programming in this domain of computations on streams.
A key innovation of ProbZelus was to demonstrate that {\em delayed sampling}~\cite{murray18delayed_sampling} could be extended to work with streams to provide high-quality inference procedures.  Delayed sampling is an inference algorithm that combines both exact and approximate inference; it takes advantage of exact inference when efficient known closed-formed solutions exist and falls back on sampling-based, approximate inference when required.
Specifically, delayed sampling combines Bayesian networks -- graphs that encode exact distributions of probabilistic models -- with particle filtering~\cite{doucet-smc-2006} -- an approximate inference algorithm.

The challenge in adapting delayed sampling to computations on streams is that such computations run for indefinite periods of time and are often subject to stringent limits on resources, such as memory. 
\citet{probzelus} showed that, in many cases, only a finite number of nodes in delayed sampling's graph data structures were reachable at any given time, and the rest could not influence the computation in the future and could be removed from memory.
However, this behavior depends on the probabilistic model under consideration; delayed sampling is not guaranteed to maintain a bounded amount of memory for all programs.
The result is then that though probabilistic programming languages are designed to hide the complexities of developing probabilistic inference algorithms, certain combinations of a model and the inference algorithm will result in undesirable behaviors that the developer did not anticipate. 
Moreover, the developer has no means to reason about these behaviors except by inspecting the implementation of the inference algorithm.

\vspace{-.15cm}
\paragraph{Bounded-Memory Delayed Sampling}
In this paper, we formalize semantic conditions under which applying delayed sampling to probabilistic programs with streams will execute in bounded memory.

The two conditions are dataflow properties of the core operations of delayed sampling: $\dsassume$, $\dsobserve$, and $\dsvalue$, which respectively add a new random variable to the delayed sampling graph, observe a random variable, and evaluate a random variable to produce a sampled value.
The \emph{$m$-consumed property} states that all variables introduced with $\dsassume$ are eventually consumed by an $\dsobserve$ or a $\dsvalue$, or are passed to other $\dsassume$s resulting in new variables that are themselves $(m-1)$-consumed. 
An {\em unseparated path} is a sequence of random variables, each passed as parameter to the $\dsassume$ operation of the next, where no variable is passed to an $\dsobserve$ or $\dsvalue$ operation. 
The \emph{unseparated paths property} states that no variable maintained in the program state starts an unseparated longer than some fixed bound~$n$. 
A program executes in bounded memory under delayed sampling if, and only if, it satisfies the $m$-consumed and unseparated paths properties.

\paragraph{Static Analysis}
We propose a static analysis that checks the $m$-consumed and unseparated paths properties to soundly ensure that any program that passes the analysis satisfies these properties, and thus executes in bounded memory under delayed sampling.

\pagebreak 
\paragraph{Contributions} In this paper, we present the following contributions:
\begin{itemize}[leftmargin=3.5mm]
    \item We introduce and formalize the $m$-consumed and unseparated paths properties, and show these are necessary and sufficient for a program to have bounded-memory execution.
    \item We present a static analysis to check these properties, and prove that the analysis is sound.
    \item We implement the analysis and evaluate it against several probabilistic inference benchmarks. Our results show that for eight of nine benchmarks, the analysis determines whether the semantic properties necessary for bounded-memory execution are satisfied, and we identify the precision limitation of conservative static analysis on the remaining benchmark.
\end{itemize}

This work brings probabilistic programming to control settings with the new benefit of static guarantees on the system's resource consumption. To the best of our knowledge, our work is the first to develop a resource analysis for a probabilistic program in relation to its probabilistic programming system's underlying inference algorithm.

The remainder of the paper is structured as follows. 
In \Cref{sec:example}, we give an example program to illustrate the concepts in the paper. 
In \Cref{sec:ideal-semantics}, we present the syntax and semantics of a language for probabilistic programming with streams, adapted from the \muF{} language from \citet{probzelus}. 
In \Cref{sec:delayed-sampling}, we review background on delayed sampling, based on the contributions from \citet{murray18delayed_sampling} and \citet{probzelus}.
In \Cref{sec:semantic-properties}, we present the $m$-consumed and unseparated paths semantic properties.
In \Cref{sec:analysis,sec:evaluation}, we present and evaluate the static analysis.
\Cref{sec:related,sec:conclusion} summarize related work and present conclusions.

\section{Example}\label{sec:example}

\Cref{fig:lqr_src} presents the example of a \zl{robot} designed to navigate to a desired position \zl{target} using measurements \zl{obs} from a noisy position sensor. 
The robot issues a command \zl{u} that indicates the acceleration to apply to change its position.
The robot (1) estimates its current position with a probabilistic model \zl{kalman} and (2) uses this estimate to compute the command \zl{u} with a deterministic \zl{controller} (e.g., a Linear-Quadratic Regulator~\citep{sontag13control}, the implementation of which we have elided for simplicity).
We present the example in $\muF$, a purely functional core calculus for probabilistic programming with streams.

\begin{wrapfigure}[16]{r}{.57\linewidth}
  \vspace{-0.25cm}
  \begin{lstlisting}[basicstyle=\footnotesize\ttfamily,aboveskip=0em,numbers=left,xleftmargin=1em]
val kalman = stream {
  init = 0.0;
  step (pre_x, obs) =
    let x = sample (gaussian (pre_x, 1.0)) in
    let () = observe (gaussian (x, 1.0), obs) in
    (x, x)
}
val robot = stream {
  init = (0.0, init controller, infer kalman);
  step ((c, k), (obs, target)) =
    let x_dist, k' = unfold (k, obs) in
    let u, c' = unfold (c, (target, mean (x_dist))) in
    (u, (c', k'))
}
\end{lstlisting}
\vspace{-.25cm}
  \caption{\muF program with main stream function \zlm{robot}.}
  \label{fig:lqr_src}
\end{wrapfigure}

The program is a set of {\em stream function} definitions that each consist of (1) an initializer, and (2)~a step function that given the previous state and an input value produces an output value and a new state~\cite{mealy-1955}.
The operators \zl{init} and \zl{infer} instantiate a stream function  by creating an internal state.
A stream function can be applied to an input stream to generate an output stream with the operator \zl{unfold}, which applies the step function using the internal state and the input values.
Unlike \zl{init}, the step function of an instance created using \zl{infer} performs probabilistic inference and thus returns at each iteration a distribution of outputs and a distribution of states.

The main stream function, \zl{robot}, has a state composed of two stream function instances: \zl{c} the deterministic \zl{controller}, and \zl{k} the \zl{kalman} probabilistic model.
The \zl{robot} initializer creates these two instances (L.9).
The transition function of instance \zl{k} performs probabilistic inference to infer a distribution of the robot's state~\zl{x_dist} and an updated instance~\zl{k'}~(L.11).
Then the transition function of instance~\zl{c} computes a command \zl{u} to go toward the destination \zl{target} using statistics of the position distribution and an updated instance \zl{c'}~(L.12).
The transition function of \zl{robot} returns the command~\zl{u} and the updated state~(L.13).

\subsection{Probabilistic Model}

The stream function \zl{kalman} specifies a \emph{hidden Markov model}~\citep{hmm}, a common probabilistic model for tracking applications in which the goal is to estimate the trajectory of an object given noisy measurements of the object's position.

The stream function's state consists of a \emph{latent} random variable, \zl{pre_x}, that denotes the position of the robot at the previous iteration.
The robot's state is latent in that the robot is unable to directly observe its position; instead it must leverage a noisy measurement or \emph{observation} of its position to infer a probability distribution over its potential states.

Inside the definition of \zl{kalman}, the program models the latent nature of \zl{x} by sampling the current position from a Gaussian distribution centered around its previous position \zl{pre_x}~(L.4).
The program models the observation by taking the observed sensor value as input, \zl{obs}, and supplying it as an input to the \zl{observe} operator.
In this example, the \zl{observe} specifies that \zl{obs} is an observation from a Gaussian distribution centered around the position \zl{x}.
The \zl{observe} operator conditions the program's execution on the observed value~(L.5) in that it adjusts the distribution that will be inferred for \zl{x}.

The sequence of diagrams in \Cref{fig:sds_kalman} illustrates the evolution of a representation of the hidden Markov model over the first four iterations of the program. Each light grey node denotes a latent random variable for \zl{pre_x} or \zl{x} at a given iteration.
Each dark grey node denotes an observation at the given iteration. Each solid black arrow signifies a dependence between random variables as in a traditional Bayesian network representation of a probabilistic graphical model~\cite{koller-graphical-models-2009}.
Of note, each observation at each iteration depends on the current position and the robot's state at a given iteration depends only on its position at the previous iteration.

\subsection{Inference with Delayed Sampling}

The \zl{kalman} probabilistic model is not sufficient for the robot to reason about its position. 
Instead, the robot must perform inference on the model to compute a posterior distribution of \zl{x} conditioned on its observations.
As mentioned, the \zl{infer} operator in the \zl{robot} stream function applies inference to the probabilistic model it receives as input.
In this paper, we study delayed sampling~\citep{murray18delayed_sampling,probzelus} as the algorithmic implementation of the \zl{infer} operator.

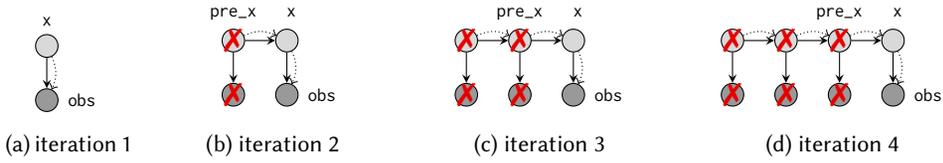
\begin{figure*}
\centering
\begin{small}
\begin{tabular}{cccc}
\begin{subfigure}[b]{.12\textwidth}
\centering
\begin{tikzpicture}[node distance=0.4cm and 0.4cm,every node/.style={font=\scriptsize, align=center}]
    \node[marginalized, label={\texttt{x}}] (x1) {};
    \node[label={[label distance=0.1ex]0:\texttt{obs}}, realized, below=of x1] (o1) {};
    \path let \p1 = (o1) in node at (\x1, \y1) {};
    \draw[-stealth] (x1) -- (o1);
    \link[->, densely dotted]{x1}{o1};
\end{tikzpicture}
\caption{iteration 1}
\label{fig:kalman-init}
\end{subfigure}
& 
\begin{subfigure}[b]{.22\textwidth}
\centering
\begin{tikzpicture}[node distance=0.4cm and 0.4cm,every node/.style={font=\scriptsize, align=center}]
    \node[marginalized, label={\texttt{pre\_x}}] (x1) {};
    \path let \p1 = (x1) in node at (\x1, \y1) {\rmark};
    \node[realized, below=of x1] (o1) {};
    \path let \p1 = (o1) in node at (\x1, \y1) {\rmark};
    \node[marginalized, label={\texttt{\phantom{p}x}}, right=of x1] (x2) {};
    \draw[-stealth] (x1) -- (x2);
    \link[->, densely dotted]{x1}{x2};
    \node[realized, label={[label distance=0.1ex]0:\texttt{obs}}, below=of x2] (o2) {};
    \path let \p2 = (o2) in node at (\x2, \y2) {};
    \link[->, densely dotted]{x2}{o2};
    \draw[-stealth] (x1) -- (o1);
    \draw[-stealth] (x2) -- (o2);
\end{tikzpicture}
\caption{iteration 2}
\label{fig:sds_kalman_addx}
\end{subfigure}
& 
\begin{subfigure}[b]{.24\textwidth}
\centering
\begin{tikzpicture}[node distance=0.4cm and 0.4cm,every node/.style={font=\scriptsize, align=center}]
    \node[marginalized] (x1) {};
    \path let \p1 = (x1) in node at (\x1, \y1) {\rmark};
    \node[realized, below=of x1] (o1) {};
    \path let \p2 = (o1) in node at (\x2, \y2) {\rmark};
    \node[marginalized, label={\texttt{pre\_x}}, right=of x1] (x2) {};
    \path let \p1 = (x2) in node at (\x1, \y1) {\rmark};
    \draw[-stealth] (x1) -- (x2);
    \link[->, densely dotted]{x1}{x2};
    \node[realized, below=of x2] (o2) {};
    \path let \p1 = (o2) in node at (\x1, \y1) {\rmark};
    \node[marginalized, label={\texttt{\phantom{p}x}}, right=of x2] (x3) {};
    \draw[-stealth] (x2) -- (x3);
    \link[->, densely dotted]{x2}{x3};
    \node[realized, label={[label distance=0.1ex]0:\texttt{obs}}, below=of x3] (o3) {};
    \path let \p3 = (o3) in node at (\x3, \y3) {};
    \link[->, densely dotted]{x3}{o3};
    \draw[-stealth] (x1) -- (o1);
    \draw[-stealth] (x2) -- (o2);
    \draw[-stealth] (x3) -- (o3);
\end{tikzpicture}
\caption{iteration 3}
\label{fig:sds_kalman_addo}
\end{subfigure}
& 
\begin{subfigure}[b]{.27\textwidth}
\centering
\begin{tikzpicture}[node distance=0.4cm and 0.4cm,every node/.style={font=\scriptsize, align=center}]
    \node[marginalized] (x1) {};
    \path let \p1 = (x1) in node at (\x1, \y1) {\rmark};
    \node[realized, below=of x1] (o1) {};
    \path let \p2 = (o1) in node at (\x2, \y2) {\rmark};
    \node[marginalized, right=of x1] (x2) {};
    \path let \p3 = (x2) in node at (\x3, \y3) {\rmark};
    \draw[-stealth] (x1) -- (x2);
    \link[->, densely dotted]{x1}{x2};
    \node[realized, below=of x2] (o2) {};
    \path let \p4 = (o2) in node at (\x4, \y4) {\rmark};
    \node[marginalized, label={\texttt{pre\_x}}, right=of x2] (x3) {};
    \path let \p1 = (x3) in node at (\x1, \y1) {\rmark};
    \draw[-stealth] (x2) -- (x3);
    \link[->, densely dotted]{x2}{x3};
    \node[realized, below=of x3] (o3) {};
    \path let \p1 = (o3) in node at (\x1, \y1) {\rmark};
    \node[marginalized, label={\texttt{\phantom{p}x}}, right=of x3] (x4) {};
    \draw[-stealth] (x3) -- (x4);
    \link[->, densely dotted]{x3}{x4};
    \node[realized, label={[label distance=0.1ex]0:\texttt{obs}}, below=of x4] (o4) {};
    \path let \p4 = (o4) in node at (\x4, \y4) {};
    \draw[-stealth] (x1) -- (o1);
    \draw[-stealth] (x2) -- (o2);
    \draw[-stealth] (x3) -- (o3);
    \draw[-stealth] (x4) -- (o4);

    \link[->, densely dotted]{x4}{o4};
\end{tikzpicture}
\caption{iteration 4}
\label{fig:sds_kalman_margx}
\end{subfigure}
\end{tabular}
\end{small}
\caption{
	The evolution of the delayed sampling graph for the hidden Markov model in \Cref{fig:lqr_src} (\zl{kalman}) as implemented by \citet{probzelus}. Each node denotes either a value (dark gray) or a distribution (light gray). A plain arrow denotes a dependency in the underlying Bayesian network. A dotted arrow denotes a pointer in the implementation of the delayed sampling graph. Each label indicates the program variable that corresponds to a node. An \rmark{} on a node denotes that the node is not reachable from the program state.
}
\label{fig:sds_kalman}
\vspace{-.25cm}
\end{figure*}

Delayed sampling is an extension of a particle filtering algorithm that leverages symbolic execution to reason about the relationship between random values and perform exact inference if possible.
A particle filter estimates the posterior distribution from a set of \emph{particles}, i.e., independent executions of the model. 
For each particle, delayed sampling operates by dynamically maintaining a graph --- i.e., a Bayesian network -- that records the dependence relationships between the random variables in the program (\Cref{fig:sds_kalman}).
The key idea is that rather than sample a concrete value for each random variable in the program (e.g., \zl{x}), delayed sampling instead returns a reference to a node in the graph.
This node contains a closed-form representation of the distribution that the \zl{sample} operator sampled from, along with the distribution's dependence on other random variables in the program.
If a symbolic computation fails, delayed sampling can fall back to a particle filter by drawing concrete values for the random variables.

\subsection{Bounded-Memory Delayed Sampling}
\label{sec:example-inference}

\newcommand{\xds}{\mathtt{x}}
\newcommand{\yds}{\mathtt{y}}

A key concern when applying delayed sampling to streams, which may execute for an indefinite number of iterations, is if the size of the delayed sampling graph is bounded from above by a fixed constant for all iterations of the program.
If not, then the delayed sampling graph may not consume bounded memory and the program may exhaust its resources if permitted to execute indefinitely. 

In general, bounding memory use is challenging because the underlying Bayesian network can in fact be unbounded.
Nevertheless, a delayed sampling implementation can maintain bounded memory for some programs, depending on the operation of said programs.
In this subsection, we review the delayed sampling implementation presented by \citet{probzelus} which can execute in bounded memory for some programs.

\paragraph{Bounded-Memory Example} \Cref{fig:sds_kalman} shows how delayed sampling maintains bounded memory for the program in \Cref{fig:lqr_src}.
For each particle, the delayed sampling implementation must keep in memory all the nodes that are \emph{reachable} from any node referenced in the program state.
The dashed lines in \Cref{fig:sds_kalman} visualize the reachability relation, where the node each line points to is reachable from the node the line points from.
As the program evolves its state and changes the variables the state contains, nodes in the delayed sampling graph may become unreachable, marked~$\rmark$.

\Cref{fig:kalman-init} shows the delayed sampling graph after the first iteration.
The graph consists of two nodes: one introduced by sampling the variable \zl{x}, and one introduced by the observation of \zl{obs}.
At the end of the step, both are in the program state and reachable.

\Cref{fig:sds_kalman_addx} shows the delayed sampling graph after the second iteration.
The program has added two nodes to the graph for sampling \zl{x} and observing \zl{obs}.
The nodes left over from the first iteration 
are still in the graph, but are no longer reachable.

\Cref{fig:sds_kalman_addo,fig:sds_kalman_margx} show the delayed sampling graph at iterations 3 and 4 respectively.
In each case, the most recently introduced nodes for \zl{x} and \zl{obs} are reachable, and the nodes from the previous iterations 
are unreachable.
In general, the program ensures that at any iteration, the most recently introduced nodes are reachable, and the rest are unreachable.
Because there are at most two reachable nodes for all iterations, inference executes in bounded memory.

\begin{figure}[t]
\begin{lstlisting}[basicstyle=\footnotesize\ttfamily,aboveskip=0em]
val kalman_first = stream {
  init = (true, 0.0, 0.0);
  step ((first, i, pre_x), obs) =
    let (i, pre_x) =
      if first then (let i = sample (gaussian (0.0, 1.0)) in (i, i))
      else (i, pre_x) in
    let x = sample (gaussian (pre_x, 1.0)) in
    let () = observe (gaussian (x, 1.0), obs) in
    (x, (false, i, x))
}
\end{lstlisting}
\vspace{-15pt}
\caption{Model with unbounded memory consumption.}
\label{fig:kalman_first_src}
\vspace{-1em}
\end{figure}

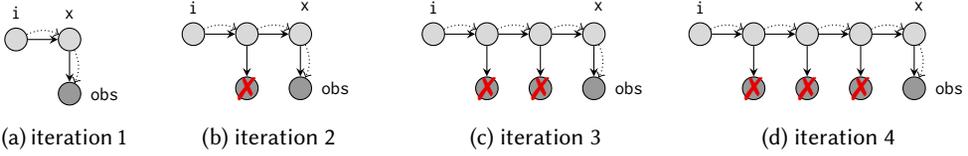
\begin{figure*}
\centering
\begin{small}
\begin{tabular}{cccc}
\begin{subfigure}[b]{.12\textwidth}
\centering
\begin{tikzpicture}[node distance=0.4cm and 0.4cm,every node/.style={font=\scriptsize, align=center}]
    \node[marginalized, label={\texttt{i}}] (i) {};
    \node[marginalized, label={\texttt{x}}, right=of i] (x1) {};
    \node[realized, label={[label distance=0.1ex]0:\texttt{obs}},  below=of x1] (o1) {};
    \path let \p1 = (o1) in node at (\x1, \y1) {};
    \draw[-stealth] (i) -- (x1);
    \draw[-stealth] (x1) -- (o1);

    \link[->, densely dotted]{i}{x1};

    \link[->, densely dotted]{x1}{o1};
\end{tikzpicture}
\caption{iteration $1$}
\label{fig:sds_kalman_first_init}
\end{subfigure}
& 
\begin{subfigure}[b]{.22\textwidth}
\centering
\begin{tikzpicture}[node distance=0.4cm and 0.4cm,every node/.style={font=\scriptsize, align=center}]
    \node[marginalized, label={\texttt{i}}] (i) {};
    \node[marginalized, right=of i] (x1) {};
    \node[realized, below=of x1] (o1) {};
    \path let \p1 = (o1) in node at (\x1, \y1) {\rmark};
    \node[marginalized, label={\texttt{\phantom{p}x}}, right=of x1] (x2) {};
    \draw[-stealth] (i) -- (x1);
    \draw[-stealth] (x1) -- (x2);
    \node[realized, label={[label distance=0.1ex]0:\texttt{obs}}, below=of x2] (o2) {};
    \path let \p2 = (o2) in node at (\x2, \y2) {};
    \draw[-stealth] (x1) -- (o1);
    \draw[-stealth] (x2) -- (o2);

    \link[->, densely dotted]{i}{x1};
    \link[->, densely dotted]{x1}{x2};

    \link[->, densely dotted]{x2}{o2};
\end{tikzpicture}
\caption{iteration 2}
\label{fig:sds_kalman_first_addx}
\end{subfigure}
& 
\begin{subfigure}[b]{.24\textwidth}
\centering
\begin{tikzpicture}[node distance=0.4cm and 0.4cm,every node/.style={font=\scriptsize, align=center}]
    \node[marginalized, label={\texttt{i}}] (i) {};
    \node[marginalized, right=of i] (x1) {};
    \node[realized, below=of x1] (o1) {};
    \path let \p1 = (o1) in node at (\x1, \y1) {\rmark};
    \node[marginalized, right=of x1] (x2) {};
    \draw[-stealth] (i) -- (x1);
    \draw[-stealth] (x1) -- (x2);
    \node[realized, below=of x2] (o2) {};
    \path let \p2 = (o2) in node at (\x2, \y2) {\rmark};
    \node[marginalized, label={\texttt{\phantom{p}x}}, right=of x2] (x3) {};
    \draw[-stealth] (x2) -- (x3);
    \node[realized, label={[label distance=0.1ex]0:\texttt{obs}}, below=of x3] (o3) {};
    \path let \p3 = (o3) in node at (\x3, \y3) {};
    \draw[-stealth] (x1) -- (o1);
    \draw[-stealth] (x2) -- (o2);
    \draw[-stealth] (x3) -- (o3);

    \link[->, densely dotted]{i}{x1};
    \link[->, densely dotted]{x1}{x2};
    \link[->, densely dotted]{x2}{x3};

    \link[->, densely dotted]{x3}{o3};
\end{tikzpicture}
\caption{iteration 3}
\label{fig:sds_kalman_first_addo}
\end{subfigure}
& 
\begin{subfigure}[b]{.27\textwidth}
\centering
\begin{tikzpicture}[node distance=0.4cm and 0.4cm,every node/.style={font=\scriptsize, align=center}]
    \node[marginalized, label={\texttt{i}}] (i) {};
    \node[marginalized, right=of i] (x1) {};
    \node[realized, below=of x1] (o1) {};
    \path let \p1 = (o1) in node at (\x1, \y1) {\rmark};
    \node[marginalized, right=of x1] (x2) {};
    \draw[-stealth] (i) -- (x1);
    \draw[-stealth] (x1) -- (x2);
    \node[realized, below=of x2] (o2) {};
    \path let \p2 = (o2) in node at (\x2, \y2) {\rmark};
    \node[marginalized, right=of x2] (x3) {};
    \draw[-stealth] (x2) -- (x3);
    \node[realized, below=of x3] (o3) {};
    \path let \p3 = (o3) in node at (\x3, \y3) {\rmark};
    \node[marginalized, label={\texttt{\phantom{p}x}}, right=of x3] (x4) {};
    \draw[-stealth] (x3) -- (x4);
    \node[realized, label={[label distance=0.1ex]0:\texttt{obs}}, below=of x4] (o4) {};
    \path let \p4 = (o4) in node at (\x4, \y4) {};
    \draw[-stealth] (x1) -- (o1);
    \draw[-stealth] (x2) -- (o2);
    \draw[-stealth] (x3) -- (o3);
    \draw[-stealth] (x4) -- (o4);

    \link[->, densely dotted]{i}{x1};
    \link[->, densely dotted]{x1}{x2};
    \link[->, densely dotted]{x2}{x3};
    \link[->, densely dotted]{x3}{x4};

    \link[->, densely dotted]{x4}{o4};

\end{tikzpicture}
\caption{iteration 4}
\label{fig:sds_kalman_first_margx}
\end{subfigure}
\end{tabular}
\end{small}
\caption{
	The evolution of the delayed sampling graph for the variant of a Kalman probabilistic model in \Cref{fig:kalman_first_src}. Nodes and edges have the same meaning as in \Cref{fig:sds_kalman}.
}
\label{fig:sds_kalman_first}
\vspace{-1em}
\end{figure*}


\paragraph{Unbounded-Memory Example}
\Cref{fig:kalman_first_src} presents an example of a program that does not execute in bounded memory.
This is a modified version of \zl{kalman} from \Cref{fig:lqr_src} that samples an initial latent position \zl{i} from a Gaussian distribution and keeps a reference to this random variable in the state.
\Cref{fig:sds_kalman_first} shows how the program in \Cref{fig:kalman_first_src} fails to maintain bounded memory.

\Cref{fig:sds_kalman_first_init} shows the delayed sampling graph after the first iteration.
The graph consists of three reachable nodes introduced by sampling the variables \zl{i} and \zl{x} and by the observation of \zl{obs}.

\Cref{fig:sds_kalman_first_addx} shows the delayed sampling graph after the second iteration.
The program has added two nodes to the graph for sampling \zl{x} and observing \zl{obs}.
Since the variable \zl{i} is in the program state, the node between \zl{i} and \zl{x} is reachable.

\Cref{fig:sds_kalman_first_addo,fig:sds_kalman_first_margx} show that in the next iterations two new nodes are introduced at each step and one remains reachable.
The primary observation to note is that the number of introduced nodes increases at every iteration. Therefore, there is no bound on the size of the delayed sampling graph and, hence, the program does not execute in bounded memory.

\subsection{Analyzing Delayed Sampling}

\newcommand{\ivar}{\mathtt{i}}
\newcommand{\xvar}{\mathtt{x}}
\newcommand{\yvar}{\mathtt{y}}
\newcommand{\pathcolor}{green!70!black}
\begin{figure}
        \begin{tikzpicture}
        \draw (0,0) node {$%
            \trace_2 =
            \xvar_1 \samplearrow \mathtt{nil} ::%
            \yvar_1 \samplearrow \xvar_1 ::%
            \traceobs \; \yvar_1 ::%
            \xvar_2 \samplearrow \xvar_1 ::%
            {\yvar_2 \samplearrow \xvar_2} ::%
            \traceobs \; \yvar_2%
        $};

        \draw (-1.85, 0.75) node(it1) {iteration 1};
        \draw[<->] ([xshift=-6.3em]it1.south) -- ([xshift=6.3em]it1.south);
        \draw (2.65, 0.75) node(it2) {iteration 2};
        \draw[<->] ([xshift=-5.9em]it2.south) -- ([xshift=5.9em]it2.south);
    \end{tikzpicture}
    \caption{A depiction of a trace of the program in \Cref{fig:lqr_src}.
        The figure depicts the trace~$\trace_2$  \Comment{and the value of the program state~$v_1$} at the end of iteration 2.
        The trace is a $::$-separated list of primitive operations, where each primitive operation is a sampling operation $\samplearrow$ or an observation operation $\traceobs$.
        In this diagram, we use $\xvar_n$ and $\yvar_n$ to refer to the random variables introduced at iteration $n$ by, respectively, sampling \zl{x} and observing \zl{obs} in \Cref{fig:lqr_src}.
        }
    \label{fig:kalman_unsep_paths}
\end{figure}
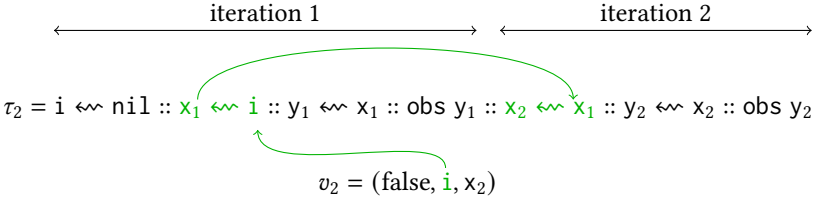
\begin{figure}
    \begin{tikzpicture}
        \draw (0,0) node {$%
            \trace_2 =
            \ivar \samplearrow \mathtt{nil} ::%
            \textcolor{\pathcolor}{\xvar_1 \samplearrow \ivar} ::%
            \yvar_1 \samplearrow \xvar_1 ::%
            \traceobs \; \yvar_1 ::%
            \textcolor{\pathcolor}{\xvar_2 \samplearrow \xvar_1} ::%
            {\yvar_2 \samplearrow \xvar_2} ::%
            \traceobs \; \yvar_2 %
        $};
        \draw (-1.9, 1.25) node(it1) {iteration 1};
        \draw[<->] ([xshift=-8em]it1.south) -- ([xshift=8em]it1.south);
        \draw (3.3, 1.25) node(it2) {iteration 2};
        \draw[<->] ([xshift=-5.9em]it2.south) -- ([xshift=5.9em]it2.south);
        \draw (0,-1) node {$v_2 = (\text{false}, \textcolor{\pathcolor}{\ivar}, {\xvar_2})$};
        \draw [\pathcolor, ->] (0.5, -0.8) .. controls (0.5, -0.25) and (-2, -1) .. (-2, -0.3);
        \draw [\pathcolor, ->] (-2.8, 0.1) .. controls (-2.8, 1) and (2.2, 1) .. (2.2, 0.1);
    \end{tikzpicture}
    \caption{
        A depiction of a trace of the program in \Cref{fig:kalman_first_src}. 
        The figure depicts the trace~$\trace_2$  and the value of the program state~$v_2$ at the end of iteration 2.
        In this diagram, we use $\ivar$, $\xvar_n$, $\yvar_n$, respectively, to refer to the random variable introduced by sampling \zl{i}, the variable introduced at iteration $n$ by sampling \zl{x}, and the variable introduced at iteration $n$ by observing \zl{obs} in \Cref{fig:kalman_first_src}.
        We have highlighted the elements of the unseparated path between $i$ and $\xvar_2$ \textcolor{\pathcolor}{in green}. }
    \label{fig:kalman_first_unsep_paths}
\end{figure}


In this paper, we present an analysis that can show that the program in \Cref{fig:lqr_src} maintains bounded memory while the program in \Cref{fig:kalman_first_src} does not.
For that, we define two dataflow properties that encode whether a program executes in bounded memory: the {\em unseparated paths} property and the {\em $m$-consumed} property.
We then show how these properties can be verified using a static analysis.

\paragraph{Traces} We formalize the dataflow properties as properties of {\em traces}.
    A trace is a recording of the important features of a program execution.
    In our case, a trace records all sampling and observation operations that the program has executed, as well as the variables that were involved in these operations.
    \Cref{fig:kalman_unsep_paths} illustrates a trace of the execution of the program in \Cref{fig:lqr_src}.

\paragraph{Unseparated Paths} An {\em unseparated path} in a trace is a sequence of variables $\{x_i\}$, where the trace specifies that each variable $x_i$ was sampled from its predecessor $x_{i-1}$ and no $x_i$ is observed.
    The {\em unseparated paths property} states that there is a uniform bound $c$ so that for all iterations no variable in the program state starts an unseparated path with more than $c$ variables in it.

    \Cref{fig:kalman_first_unsep_paths} illustrates the trace for the program in \Cref{fig:kalman_first_src}.
    This program carries the variable $\ivar$ in the program state, and because the trace specifies that $\xvar_1$ was sampled from $\ivar$, and $\xvar_2$ was sampled from $\xvar_1$, the sequence $\ivar, \xvar_1, \xvar_2$ is an unseparated path with 3 variables.
    In general, at iteration $n$, the program in \Cref{fig:kalman_first_src} maintains that $\ivar$ is in the program state and starts an unseparated path with length $n+1$.
    Because no bound can exist on the length of this path for an arbitrary number of iterations, this program fails the unseparated path property.

\paragraph{$m$-consumed} A variable is \emph{$m$-consumed} if it is no more than $m$ sampling operations away from a variable that is {\em consumed} by an observe statement. 
The {\em $m$-consumed property} states that there is a uniform bound $\overline{m}$ such every variable introduced by a sampling operation is $m$-consumed for some $m \le \overline{m}$.
We note that the traces in \Cref{fig:kalman_unsep_paths,fig:kalman_first_unsep_paths} satisfy the $m$-consumed property, because every variable is at most $2$-consumed.
For all $t$, $\yvar_t$ is $0$-consumed because it is directly observed, and $\xvar_t$ is $1$-consumed because $\yvar_t$ is sampled from $\xvar_t$ and $\yvar_t$ is $0$-consumed. 
The variable $\ivar$ is $2$-consumed because  $\xvar_1$ is sampled from $\ivar$, and $\xvar_1$ is $1$-consumed.

The \emph{Outlier} benchmark presented in Section 7 is an example of a program that fails the $m$-consumed property, and thus does not execute in bounded memory. This program sometimes observes values close to the true latent state but otherwise observes values from an outlier distribution. When the program observes a value from the outlier distribution, it fails to observe any dependencies of the latent state, and thus cannot guarantee that the latent state is $m$-consumed. Over time, if the program performs latent state updates that remain unobserved (due to the program always observing from the outlier distribution), the lack of this guarantee results in there being no uniform bound $\overline{m}$ under which the latent state could be $\overline{m}$-consumed.

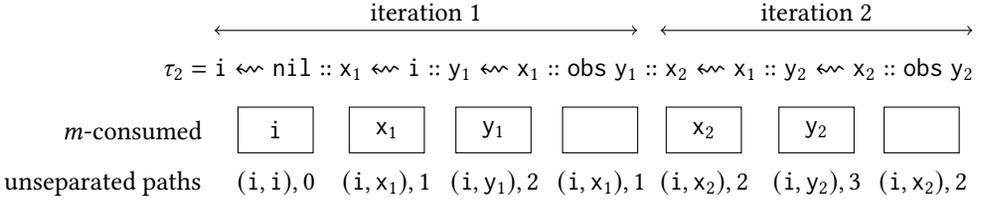
\begin{figure}
    \begin{tikzpicture}
        \draw (0,0) node {$%
            \trace_2 =
            \ivar \samplearrow \mathtt{nil} ::%
            \xvar_1 \samplearrow \ivar ::%
            \yvar_1 \samplearrow \xvar_1 ::%
            \traceobs \; \yvar_1 ::%
            \xvar_2 \samplearrow \xvar_1 ::%
            \yvar_2 \samplearrow \xvar_2 ::%
            \traceobs \; \yvar_2 %
        $};
        \draw (-1.9, 0.75) node(it1) {iteration 1};
        \draw[<->] ([xshift=-8em]it1.south) -- ([xshift=8em]it1.south);
        \draw (3.3, 0.75) node(it2) {iteration 2};
        \draw[<->] ([xshift=-5.9em]it2.south) -- ([xshift=5.9em]it2.south);

        \draw (-5.8,-0.8) node {$m$-consumed};
        \draw (-4.4,-1.1) rectangle (-3.4,-0.5) node[pos=.5] {$\ivar$};
        \draw (-2.92,-1.1) rectangle (-1.92,-0.5) node[pos=.5] {$\xvar_1$};
        \draw (-1.5,-1.1) rectangle (-0.5,-0.5) node[pos=.5] {$\yvar_1$};
        \draw (-0.08,-1.1) rectangle (0.92,-0.5) node[pos=.5] {};
        \draw (1.3,-1.1) rectangle (2.3,-0.5) node[pos=.5] {$\xvar_2$};
        \draw (2.8,-1.1) rectangle (3.8,-0.5) node[pos=.5] {$\yvar_2$};
        \draw (4.2,-1.1) rectangle (5.2,-0.5) node[pos=.5] {};

        \draw (-6.2,-1.5) node {unseparated paths};
        \draw (-3.9,-1.5) node {$(\ivar, \ivar), 0$};
        \draw (-2.43,-1.5) node {$(\ivar, \xvar_1), 1$};
        \draw (-0.99,-1.5) node {$(\ivar, \yvar_1), 2$};
        \draw (0.43,-1.5) node {$(\ivar, \xvar_1), 1$};
        \draw (1.8,-1.5) node {$(\ivar, \xvar_2), 2$};
        \draw (3.3,-1.5) node {$(\ivar, \yvar_2), 3$};
        \draw (4.71,-1.5) node {$(\ivar, \xvar_2), 2$};
    \end{tikzpicture}
    \caption{
        A depiction of the abstract graphs of the program in \Cref{fig:kalman_first_src}, with the same trace as \Cref{fig:kalman_first_unsep_paths}.
        At each operation, we depict the $m$-consumed abstract graph, a set of nodes that have been introduced but not consumed.
        Because this set is empty at the end of any iteration, the program satisfies the $m$-consumed semantic property.
        The unseparated paths abstract graph is a mapping, for each unseparated path in the graph, from its endpoints to its length. We depict the longest path in the mapping.
        After each iteration, this longest path continues to lengthen, so the program does not satisfy the unseparated paths semantic property.
    }
    \label{fig:kalman_first_abstract_graphs}
\end{figure}

\paragraph{Analysis}
Our goal is ultimately to analyze whether a given program executes in bounded memory.
As we show in \Cref{sec:semantic-properties}, a program execution maintains bounded memory if and only if it satisfies both the unseparated path and $m$-consumed properties.
This reduces the problem of analyzing the bounded-memory behavior of a program to analyzing these dataflow properties. Our analysis utilizes an abstract delayed sampling graph, formally defined in \Cref{sec:analysis}, with the key aspects of these properties. For $m$-consumed, the abstract graph maintains a set of variables that have been introduced but not yet consumed, and for unseparated paths, it maintains an upper bound on their length. For example, the abstract graphs for the trace in \Cref{fig:kalman_first_unsep_paths} are given in \Cref{fig:kalman_first_abstract_graphs}.

\section{Language Model}
\label{sec:ideal-semantics}

In this section, we present a semantics for probabilistic programs with streams using the language \muF{}. We have adapted \muF{} from \citet{probzelus}'s core calculus for probabilistic programs and extended it with syntax for explicit streams.

\subsection{Syntax}

The syntax of the \muF language is defined according to the following grammar:
$$
\begin{array}{rc@{~}l}
  \mathit{program} & ::= & \phantom{\mid~}
     d^*\;m
\\[0.5em]
  d & ::= & \phantom{\mid~}
                 \muFval{p}{e}
            \mid \muFfun{f}{p}{e}
            \mid \muFval{m}{\muFstream{e}{p}{p}{e}}
\\[0.5em]
  e & ::= & \phantom{\mid~}
                 v
            \mid \muFop{v}
            \mid \muFapp{f}{v}
            \mid \muFif{v}{e}{e}
            \mid \muFletin{p}{e}{e}
\\&&
            \mid \muFinit{m}
            \mid \muFunfold{x}{v}
            \mid \muFsample{v}
            \mid \muFobserve{v}{v}
            \mid \muFinfer{m}
\\[0.5em]
  v & ::= & \phantom{\mid~}
                 c
            \mid x
            \mid \muFpair{v}{v}
\\[0.5em]
  p & ::= &  \phantom{\mid~} x \mid \muFpair{p}{p}
\end{array}
$$
A program is a set of value, function, and stream function definitions followed by the name of the main stream function.
A \emph{stream function} $m$ is composed of an initial state (\zl{init}) and a transition function (\zl{step}).
Given a state and an input, the transition function returns an output and a new state.
An expression is either a value (constant, variable, or pair), the application of a primitive operator (arithmetic operator, distribution, etc.), a function call, a conditional, or a local definition.

The expression $\muFinit{m}$ creates an \emph{instance} of a stream function, and $\muFunfold{x}{v}$ applies the instance $x$ of a stream function on an input and returns the next element and the updated instance.
Finally, the set of expressions comprises the probabilistic operators \zl{sample}, \zl{observe}, and \zl{infer}.
Nested inference and higher-order functions on streams are not allowed in the language.
We require that arguments for all syntactic operators are values to simplify the presentation of the semantics.
Since new variables can always be introduced to capture the value of any expression, this choice does not reduce the expressiveness of the language.

\subsection{Semantics}

\begin{figure}
$$
\begin{array}{@{}l@{\;\;}c@{\;\;}l}
\sem{\muFval{x}{e}}_\gamma &=& \gamma[x \leftarrow \sem{e}_\gamma]
\\[0.5em]

\sem{\muFfun{f}{p}{e}}_\gamma &=& \gamma[f \leftarrow (\fun{v}\sem{e}_{\gamma+[v/p]})]
\\[0.5em]

\multicolumn{3}{@{}l}{
\sem{\muFval{m}{\muFstream{e_{\textit{init}}}{p_{\mathit{state}}}{p_{\mathit{input}}}{e}}}_\gamma
}
\\&=& \gamma[m \leftarrow \muFstream{e_{\textit{init}}}{p_{\mathit{state}}}{p_{\mathit{input}}}{e}_\gamma]
\\[1em]

\sem{\muFinit{m}}_\gamma &=&
  \letin{\muFstream{e_{\mathit{init}}}{p_{\mathit{state}}}{p_{\mathit{input}}}{e}_{\gamma'} = \sem{m}_\gamma} \\&&
  \letin{s_{\mathit{init}} = \sem{e_{\mathit{init}}}_{\gamma'}}\,
  (s_{\mathit{init}}, \fun{(s, v)} \sem{e}_{\gamma'+[s/p_{\mathit{state}}, v/p_{\mathit{input}}]})
  \\ & & \text{if $e$ is deterministic}
  \\[0.5em]

\sem{\muFinit{m}}_\gamma &=&
  \letin{\muFstream{e_{\mathit{init}}}{p_{\mathit{state}}}{p_{\mathit{input}}}{e}_{\gamma'} = \sem{m}_\gamma} \\&&
  \letin{s_{\mathit{init}} = \sem{e_{\mathit{init}}}_{\gamma'}}\,
  (s_{\mathit{init}}, \fun{(s, v)} \psem{e}_{\gamma'+[s/p_{\mathit{state}}, v/p_{\mathit{input}}]})
  \\ & & \text{if $e$ is probabilistic}
  \\[0.5em]

\sem{\muFunfold{x}{v}}_\gamma &=&
  \begin{array}[t]{@{}l}
    \letin{v_{\mathit{state}}, f = \sem{x}_\gamma}\\
    \letin{v_{\mathit{output}}, v_{\mathit{state}}' = f(v_{\mathit{state}}, \sem{v}_\gamma)} \\
    (v_{\mathit{output}}, (v_{\mathit{state}}', f))
  \end{array}
\\[3.em]

\sem{\muFinfer{m}}_\gamma &=&
  \letin{\muFstream{e_{\mathit{init}}}{p_{\mathit{state}}}{p_{\mathit{input}}}{e}_{\gamma'} = \sem{m}_\gamma} \\&&
  \letin{s_{\mathit{init}} = \sem{e_{\mathit{init}}}_{\gamma'}}\,
  (\delta_{s_{\mathit{init}}}, \mathit{infer}(\fun{(s,v)} \psem{e}_{\gamma'+[s/p_{\mathit{state}}, v/p_{\mathit{input}}]})) \\ &&
  \text{where }
  \begin{array}[t]{@{}l}
    \mathit{infer}(f) = \fun{(\sigma, v)}
      \begin{array}[t]{@{}l}
      \letin{\mu = \fun{U} \int_{S} \sigma(ds) f(s, v)(U)}\\
      \letin{\nu = \fun{U} \mu(U) / \mu(\top)}\\
      (\pi_{1*}(\nu), \pi_{2*}(\nu))\\[1em]
    \end{array}
  \end{array}
\\
\end{array}
$$
\vspace{-20pt}
\caption{Deterministic semantics of \muF (complete definition in \Cref{fig:muF-sem-full}). }
\label{fig:muF-sem}
\end{figure}

The execution of a program $p = d^* m$ comprises three steps.
First, declarations~$d^*$ are evaluated to produce an environment $\gamma$ which contains the definition of the main stream function~$m$.
Second, an instance of the stream function $m$ is created.

Third, the instance is iteratively applied on an input stream~$(i_n)_{n \in \mathbb{N}}$ to produce an output stream~$(o_n)_{n \in \mathbb{N}}$, defined in the following way:
$$
\begin{array}{l@{~}c@{~}ll}
\sem{p}(i)_{n} & = & o_n
& \text{ where }
     \begin{array}[t]{r@{~}c@{~}l@{\qquad}r@{~}c@{~}ll}
     p & = & d^* m &
     \gamma & = & \sem{d^*}_\emptyset \\
     s_{0} & = & \sem{\muFinit{m}}_\gamma &
     o_{n}, s_{n+1} & = & \sem{\muFunfold{s_{n}}{i_n}}_\gamma & \forall n \geq 0 \\
     \end{array}
\end{array}
$$

\Cref{fig:muF-sem} defines the semantics of declarations and deterministic expressions $\sem{\cdot}$.
The declarations build the evaluation environment~$\gamma$ which maps names to values, functions, and stream functions.

The semantics of deterministic expressions corresponds to a first order functional language with new constructs to handle streams and the $\muFinfer{\cdot}$ operator~(the complete definition is given in \Cref{fig:muF-sem-full} of \Cref{sec:ideal-semantics-app}).
The expression $\muFinit{m}$ creates an instance of the stream function~$m$: a pair corresponding to the current state, and the transition function.
The current state is initialized with the value of the \zl{init} field.
The expression $\muFunfold{x}{v}$ executes the transition function of the instance~$x$ on its current state and the input~$v$. This expression produces a pair composed of the transformed value and the updated instance.

The ideal semantics of \muF probabilistic expressions $\psem{\cdot}$ is a measure-based semantics similar to the one presented by \citet{staton17} (the complete definition is given in \Cref{sec:ideal-semantics-app}).
Given an environment $\gamma$, an expression is interpreted as a measure $\psem{e}_{\gamma}: \Sigma_D \rightarrow [0, \infty)$, that is, a function which associates a positive number to each measurable set $U \in \Sigma_D$, where $\Sigma_D$ denotes the $\Sigma$-algebra of the domain of the expression $D$ (i.e., the set of measurable sets of possible values).
$\muFsample{v}$ returns the distribution $\sem{v}_{\gamma}$.
$\muFobserve{v_1}{v_2}$ weights execution paths using the likelihood of the observation~$\sem{v_2}_{\gamma}$ w.r.t. the distribution $\sem{v_1}_{\gamma}$ (for a distribution $\mu$ we denote its probability density function as $\mu_{\rm{pdf}}$).
Local definitions are interpreted as integration, and we use the Dirac delta measure to interpret deterministic expressions.

The $\muFinfer{m}$ operator creates an instance of a probabilistic stream:
the initial state is a Dirac delta distribution on the initial state of~$m$, and the transition function is $\mathit{infer}(f)$ where $f$ is the transition function of~$m$.
The body of~$f$~(the expression~$e$) is interpreted with the probabilistic semantics which defines a measure over pairs of output values and states.
The function $\mathit{infer}(f)$ takes as arguments a distribution of states~$\sigma$ and an input~$v$ and returns a distribution of outputs and a distribution of new states.
These two distributions are obtained by integrating the transition function~$f$ along the distribution~$\sigma$ of possible states (domain $S$) to build a measure~$\mu$ which is then normalized to build a distribution~$\nu$ of pairs (outputs, states).
The distribution~$\nu$ is then split into a pair of marginal distributions using the pushforward of~$\nu$ across the projections~$\pi_1$ and~$\pi_2$.

\section{Delayed Sampling}
\label{sec:delayed-sampling}

\begin{figure}
$$
\begin{array}{@{}l@{\;\;}c@{\;\;}l}

\psem{v}_{\gamma} &=& \fun{g, w}
  (\sem{v}_\gamma, g, w)
\\[0.5em]

\psem{\muFop{v}}_{\gamma} &=&
    \fun{g, w}
    (\dsapp{\op}{\sem{v}_\gamma}, g, w)
\\[0.5em]

\psem{\muFapp{f}{v}}_{\gamma} &=&
    \fun{g, w}
    \gamma(f)(\sem{v}_\gamma)(g, w)
\\[0.5em]

\psem{\muFletin{p}{e_1}{e_2}}_{\gamma} & = &
  \fun{g, w}
  \begin{array}[t]{@{}l@{}}
  \letin{v_1, g_1, w_1 = \psem{e_1}_{\gamma}(g, w)}
  \psem{e_2}_{\gamma+[v_1/p]}(g_1, w_1) \\[0.5em]
  \end{array}
\\

\psem{\muFif{v}{e_1}{e_2}}_{\gamma}  & = &
  \fun{g, w}
  \begin{array}[t]{@{}l@{}}
    \letin{b, g_b = \dsvalue(\sem{v}_{\gamma}, g)}\\
    \mathit{if} \; b \; \mathit{then} \;
      \psem{e_1}_{\gamma}(g_b, w) \;
    \mathit{else} \;
      \psem{e_2}_{\gamma}(g_b, w) \\[0.5em]
  \end{array}
\\[0.5em]

\psem{\muFunfold{x}{v}}_\gamma &=&
  \fun{g, w}
  \begin{array}[t]{@{}l}
    \letin{v_{\mathit{state}}, f = \sem{x}_\gamma}\\
    \letin{(v_{\mathit{output}}, v_{\mathit{state}}'), g' w' = f(v_{\mathit{state}}, \sem{v}_\gamma)(g, w)} \\
    ((v_{\mathit{output}}, (v_{\mathit{state}}', f)), g' w')
  \end{array}
\\[3.em]

\psem{\muFsample{v}}_{\gamma} & = &
  \fun{g, w}
  \begin{array}[t]{@{}l@{}}
    \letin{X, g' = \dsassume (\sem{v}_{\gamma}, g)}
    (X, g', w)
  \end{array}
\\[0.5em]

  \psem{\muFobserve{v_1}{v_2}}_{\gamma} & = &
  \fun{g, w}
  \begin{array}[t]{@{}l@{}}
    \letin{X, g_x = \dsassume(\sem{v_1}, g)}\\
    \letin{v, g_v = \dsvalue(\sem{v_2}, g_x)}\\
    \letin{g' = \dsobserve(X, v, g_v)}
    (\textrm{()}, g', w * \mu_{\rm pdf}(v))
  \end{array}
\\[0.5em]
\end{array}
$$
\vspace{-10pt}
\caption{Delayed sampling semantics. Probabilistic expressions are functions from a graph and a weight to a triplet (value, graph, weight).}
\label{fig:ds-sem}
\end{figure}

In this section, we present the details of delayed sampling that underpin this work.
This is a new formalization of results that were presented by \citet{murray18delayed_sampling} and \citet{probzelus}.

Delayed sampling is a semi-symbolic algorithm combining exact inference and -- when exact computation fails -- approximate inference with particle filtering~\citep{doucet-smc-2006}.
A particle filter launches multiple executions of the model.
Each execution --- or particle --- is associated to a weight.
In the operational semantics, \muFsample{d} statements draw samples from the corresponding distributions, and \muFobserve{x}{d} statements update the weight to reflect the quality of the samples.
At the end of the executions the results of all the particles are normalized according to their weights to form a categorical distribution that approximates the posterior distribution of the model.

In delayed sampling, each particle contains a graph of random variables and their dependencies that can be used to compute closed-form distributions.
Observations can be incorporated by analytically \emph{conditioning} the network.
If symbolic conditioning fails, inference falls back to a particle filter, drawing concrete samples for required random variables.

\subsection{Operational Semantics}
\label{sec:ds_op_sem}

The definition of \zl{infer} in \Cref{fig:muF-sem} makes use of an intractable integral.
The delayed sampling semantics replaces this integral by a discrete sum over the set of particles of the particle filter.
Compared to traditional particle filtering, delayed sampling performs exact computations when possible. 
Thus, we extend values $v$ with symbolic terms.
Symbolic terms include random variables~($X$) --- the nodes of the delayed sampling graph --- and applications of operators.
\[
\begin{array}{rcl}
  v & ::= & ...
            \mid X
            \mid \dsapp{\op}{v}
\end{array}
\]

The semantics in \Cref{fig:ds-sem} rely on the following high-level operations to update the graph~$g$.
\begin{description}
  \item[$v', g' = \dsvalue(v, g)$] samples all the random variables in~$v$ to produce a concrete value.
  \item[$g' = \dsobserve(X, v, g)$] 
  conditions the graph on the fact that the random variable $X$ takes the value~$v$. 
  \item[$X, g' = \dsassume(d, g)$] adds and returns a new random variable $X$ with distribution~$d$.
\end{description}

\paragraph{Probabilistic semantics}
The semantics of a probabilistic expressions are defined in \Cref{fig:ds-sem}.
The semantics of an expression $\psem{e}_{\gamma, g, w}$ takes two additional arguments: $g$, the delayed sampling graph, and $w$, the weight for the particle filter, and returns a symbolic value, an updated graph, and an updated weight.
Operator application $\muFop{v}$ introduces a symbolic expression $\dsapp{\op}{v}$.
\zl{if} uses the $\dsvalue$ operation to sample a concrete value for the condition.
$\muFsample{v}$ introduces a new random variable in the graph with distribution $v$.
$\muFobserve{v_1}{v_2}$ introduces a fresh random variable~$X$ with distribution $v_1$, and conditions the graph on the fact that $X$ takes the value~$v_2$.

\paragraph{Inference}
Given a transition function $f$, a distribution over states $\sigma$ from the previous iteration, and inputs $v_i$, the \zl{infer} operator computes a distribution of outputs and new distribution over states for the next iteration.
First, the inference draws~$N$ states from $\sigma$.
Each of theses states~$s_n$ is associated with a delayed sampling graph~$g_n$.
Second, the transition function~$f$ returns a symbolic output value $v_n$, a new state $s_n'$, the updated graph $g_n'$, and the importance weight $w_n$.
Third, the $\dsdistribution(o_n, g_n')$ function
returns a distribution of values without altering the graph, and the new distribution over states is a Dirac delta distribution on the pair $(s_n', g_n')$.
Finally, results are accumulated in a mixture distribution using the weights $w_n$ and this distribution is split into a distribution of values and a distribution of next states.\footnote{we write $\overline{w_i} = w_i / \sum_{i=1}^N w_i$ for the normalized weights.}
$$
\begin{array}{l}
\sem{\muFinfer{m}}_\gamma ~=~
  \begin{array}[t]{@{}l}
  \letin{\muFstream{e_{\mathit{init}}}{p_{\mathit{state}}}{p_{\mathit{input}}}{e}_{\gamma'} = \sem{m}_\gamma} \\
  \letin{s_{\mathit{init}} = \sem{e_{\mathit{init}}}_{\gamma'}}\,
  (\delta_{(s_{\mathit{init}}, \emptyset)}, \mathit{infer}(\fun{(p_{\mathit{state}}, p_{\mathit{input}})} \psem{e}_{\gamma'}))
  \end{array}
  \\ \quad
  \text{where }
  \begin{array}[t]{@{}l}
    \mathit{infer}(f) = \fun{(\sigma, v_i)}

    \begin{array}[t]{@{}l@{}}
    \letin{\mu =
      \begin{array}[t]{@{}l@{}}
      \fun{U} \sum\limits_{n=1}^N \;
        \begin{array}[t]{@{}l@{}}
          \letin{s_n, g_n = \dsdraw(\sigma)}\\
          \letin{(o_n, s_n'), g_n', w_n = f{\muFpair{s_n}{v_i}}(g_n, 1)}\\
          \letin{d_n = \dsdistribution(o_n, g_n')}\\
          \overline{w_n} *  d_n(\pi_1(U)) * \delta_{s_n', g_n'}(\pi_2(U))
        \end{array}
      \end{array}\\}
      (\pi_{1*}(\mu), \pi_{2*}(\mu))
    \end{array}
  \end{array}
\end{array}
$$

\subsection{Graph Manipulation}

We now describe the graph manipulation functions that are required to define the high-level operations $\dsvalue$, $\dsassume$, and $\dsobserve$ used in the semantics of \Cref{fig:ds-sem}.
\citet{lunden17} and \citet{murray18delayed_sampling} provide detailed explanations of these operations.

\paragraph{Notation} In this section and those that follow, $\freerandvars(\progvar)$ denotes the {\em free random variables} of a program value $\progvar$, i.e., the set of variables used in the symbolic expression $\progvar$.

\paragraph{Graph Data Structure}
A delayed sampling graph~$g$ is defined by a tuple $(V, E, q)$ where
$V$ is a set of vertices -- the random variables,
$E$ is a set of directed edges -- the dependencies between random variables, and
$q$ is a relation mapping each node to a state: $\stinitialized$, $\stmarginalized$, or $\strealized$.

A node $\stinitialized(p_{X|Y})$ represents a random variable~$X$ with a conditional distribution~$p_{X|Y}$ where~$Y$ is the unique parent of~$X$.
A node $\stmarginalized(p_X)$ represents a random variable~$X$ with a marginal distribution~$p_X$.
A $\stmarginalized$ node has at most one parent. 
If there is a parent node, the distribution~$p_X$ incorporates its distribution.
A node $\strealized(v)$ represents a random variable~$X$ associated to a concrete value~$v$.
By construction, a delayed sampling graph is a forest -- a set of trees~(each node has at most one parent).

\paragraph{$\dsvalue$}
The operation $\dsvalue(v, g)$ converts the symbolic expression~$v$ into a concrete value by sampling all the random variables in~$v$.
All these random variables become $\strealized$ nodes in the graph, and the distributions depending on these variables are updated.
$$
\begin{array}{lcl}
  \dsvalue(v, g) & = & (v, g) \quad \text{ if $v$ is a concrete value} \\
  \dsvalue(\dsapp{\mathit{op}}{v}, g) & = &
    \letin{v', g' = \dsvalue(v, g)} (\mathit{op}(v'), g') \\
  \dsvalue(X, g) & = &
     \letin{V, E, q = g} \\&&
     \mathit{if}\ q(X) = \strealized(v)\ \mathit{then}\ (v, g) \\&&
     \mathit{else}\ 
       \begin{array}[t]{@{}l}
         \letin{V', E', q'[ X \leftarrow \stmarginalized(\mu)] = \dsllgraft(X, g)} \\
         \letin{v = \dsdraw(\mu)} \\
         (v, (V', E', q'[X \leftarrow \strealized(v)]))
       \end{array}
\end{array}
$$

\noindent
If~$v$ is already a concrete value, there is nothing to do.
If~$v$ is the application of an operator, $\dsvalue$ recursively samples a concrete value for the argument and applies the operator to this value.
If~$v$ is a random variable~$X$ that is already realized, $\dsvalue$ returns the corresponding value.
Otherwise, $\dsvalue$  (1) calls the $\dsllgraft$ function defined in~\Cref{sec:graft-app} to marginalize~$X$ and all its ancestors, (2) draws a sample from the marginalized distribution, and (3) returns this value and turns $X$ into a $\strealized$ node.
Note that $\dsllgraft$ might have to realize some nodes since it marginalizes all its ancestors and a marginal node has a single marginalized child.
During marginalization, $\dsllgraft$ also removes edges between $\stmarginalized$ nodes and their $\strealized$ child if any.

\paragraph{$\dsassume$}
The operation $\dsassume(v, g)$ adds a new random variable~$X$ with distribution~$v$ in graph~$g$.
$$
\begin{array}{lcl}
  \dsassume(v, g) & = &
    \letin{(V, E, q) = g} \\&&
    \letin{X = \fresh(V)} \\&&
    \mathit{if}\ \freerandvars(v) = \emptyset\ \mathit{then}\ 
      (X, (V \cup \{ X \}, E, q[X \leftarrow \stmarginalized(v)])) \\&&
    \mathit{else}\ \mathit{if}\ \freerandvars(v) = \{ Y \}\ \wedge \conjugate(v, Y, g)\ \mathit{then} \\&&
                                                                                                 \quad (X, (V \cup \{ X \}, E \cup \{(X, Y)\}, q[X \leftarrow \stinitialized(v)])) \\&&
    \mathit{else} \\&&
      \quad \letin{v', (V,' E', q') = \dsvalue(v, (V \cup \{ X \}, E, q))}\\&&
      \quad (X, (V', E', q'[X \leftarrow \stmarginalized(v')]))
\end{array}
$$
The distribution~$v$ is a symbolic expression which can be a marginal distribution that does not depends on other random variables --- e.g.,
$\dsapp{\mathit{bernoulli}}{0.5}$--- or a conditional distribution --- e.g., $\dsapp{\mathit{bernoulli}}{Y}$ where $Y$ is a random variable.
If~$v$ is a marginal distribution, $\dsassume$ just adds a new marginalized node in the graph.
If~$v$ is a conditional distribution, $\dsassume$ tries to keep track of the dependency between $X$ and a random variable used in $v$ (the delayed sampling graph is a forest where each node has at most one parent).

The value $v$ thus represents a distribution~$p_{X|Y}$ where~$X$ depends on a unique random variable~$Y$.
If the distribution $p_{X|Y}$ and $p_Y$ are conjugate~(${\conjugate(v, Y, g)}$) --- e.g., $\dsapp{\mathit{bernoulli}}{Y}$ with $Y \sim \mathit{beta}(\alpha, \beta)$ --- marginalization and conditioning are tractable operations, and $\dsassume$ adds an edge between $Y$ and a new initialized node $X$ to the graph.
Otherwise, symbolic computation is not possible; $\dsassume$ calls $\dsvalue$ to sample a concrete value, thus breaking the dependency, and adds a new independent $\stmarginalized$ node to the graph.

\paragraph{$\dsobserve$}
The operation $\dsobserve(X, v, g)$ assigns the concrete value~$v$ to~$X$ and updates the distributions depending on~$X$ accordingly.
$$
\begin{array}{lcl}
  \dsobserve(X, v, g) & = &
    \letin{(V, E, q) = \dsllgraft(X, g)} 
    (V, E, q[X \leftarrow \strealized(v)])
\end{array}
$$
Similarly to $\dsvalue$, the $\dsobserve$ operation uses the function $\dsllgraft$ to marginalize the variable~$X$ and then turns~$X$ to a $\strealized$ node associated with the value~$v$.

\subsection{Memory Usage}
\label{subsec:ds_mem_usage}

\citet{probzelus} proposed an implementation of delayed sampling where an $\stinitialized$ node only has a pointer to its parent, a $\stmarginalized$ node only has a pointer to its unique $\stmarginalized$ or $\strealized$ child, if any, and a $\strealized$ node has no pointers to its parent or any of its children.

\paragraph{Garbage Collection}
A node in the delayed sampling graph can be safely removed if none of the program variables depend on its value.
We assume the existence of a garbage collection routine that deallocates the nodes of the graph that are not {\em reachable} as soon as possible.

\begin{definition}[Reachability] Given a set of root variables $r$ and a delayed sampling graph $g =(V, E, q)$, the set of {\em reachable variables} -- written $\reachable(g, r)$ -- is defined as follows:
    $$
    \begin{array}{rcl}
      R & = & \{ (X, Y) \mid \big((X, Y) \in E \wedge q(X) = \stinitialized \big) \vee%
        \big((Y, X) \in E \\
        && \quad \wedge \; q(X) = \stmarginalized \wedge (q(Y) = \stmarginalized \vee q(Y) = \strealized)\big)\\
    \reachable(g, r) &=& \{Y \mid  (R^*(X, Y)) \wedge X \in r \wedge Y \in V \}
    \end{array} 
    $$
where $R^*$ denotes the reflexive transitive closure of the relation $R$.
\end{definition}

If we consider the graph in \Cref{fig:sds_kalman_addx}, $\reachable(g, \{ \texttt{x} \}) = \{ \texttt{x} \}$.
In the example of \Cref{fig:sds_kalman_first_addx}, we have $\reachable(g, \{ \texttt{i}, \texttt{x} \}) = \{ \texttt{i}, \texttt{pre\_x}, \texttt{x}  \}$, where \texttt{pre\_x} is the gray node in between the nodes for \texttt{i} and \texttt{x}.
Reachability is the core property used in Definition~\ref{def:ll-streaming} to define what it means for a program to run in bounded memory.

\paragraph{Graph Expansion}

The only operation that increases the size of the graph is $\dsassume$ which introduces new nodes.
The operations $\dsvalue$ and $\dsobserve$ can only marginalize and realize nodes.
If~$g'$ is the graph resulting from the application of $\dsvalue$ or $\dsobserve$ on a graph~$g$, $g$ and $g'$ have the same structure but  $\stinitialized$ nodes can be $\stmarginalized$ or $\strealized$, and $\stmarginalized$ nodes can be $\strealized$.
The reachability relation of the graph implies that $\dsvalue$ and $\dsobserve$ reduce the number of dependencies in the delayed sampling graph, that is, ${\reachable(g', r) \subseteq \reachable(g, r)}$.

\paragraph{Initialized and Marginalized Chains}
Two patterns can yield unbounded memory consumption.
First, it is possible to keep adding nodes without realizing them (via observation or sampling), thus forming \emph{initialized chains}.
An initialized chain is a sequence of initialized nodes, each of which holds a pointer to its parent and thereby expands the number of random variables that are reachable.
Second, it is possible that nodes are only indirectly used to realize one of their children.
These marginalized nodes can form \emph{marginalized chains}.
A marginalized chain is a sequence of marginalized nodes, each of which holds a pointer to its child and thus expands the number of random variables that are reachable.
The last node of a marginalized chain may be realized.

\section{Semantic Properties}
\label{sec:semantic-properties}

\newcommand{\notn}{c}

In this section, we define conditions under which delayed sampling executes in bounded memory.
We define these conditions as properties of {\em executions}.
An execution is a sequence of pairs of a state and a delayed sampling graph $(s_n, g_n)_{n \in \mathbb{N}}$, where each state is a semi-symbolic value as defined in \Cref{sec:ds_op_sem}.
An execution defines the sequence of states and graphs a model --- i.e., an argument of an \zl{infer} --- goes through.

The inference step function $\mathit{infer}(f)$ in $\sem{\zl{infer(}m\zl{)}}$ may operate over multiple executions of~$f$~(see \Cref{sec:ds_op_sem}).
However, $\mathit{infer}(f)$ executes in bounded memory if every execution of $f$ is bounded-memory.
This is because $\mathit{infer}(f)$ always updates its state by mapping $f$ over states and graphs from the distribution at the previous iteration.
Thus, any state and graph in the distribution at the next iteration must have come from {\em some} execution of $f$, and if all executions of $f$ are bounded-memory, all states and graphs in the distribution must have bounded memory.
We have formalized this in more details in \Cref{subsec:execution}.

\medskip

Based on this notion of execution, we introduce two notions of bounded-memory executions of delayed sampling, and semantic properties which are necessary and sufficient for bounded-memory execution.
In Section~\ref{sec:llstreaming} we present a low-level definition of bounded memory that directly corresponds to how the delayed sampling runtime executes.
In Section~\ref{sec:hlstreaming} we present an alternative high-level definition in terms of dataflow properties of the high-level delayed sampling operators: the \emph{$m$-consumed} and \emph{unseparated paths} properties.
In Section~\ref{sec:ll-hl-equiv} we show that the high-level and low-level formulations are equivalent.
In particular, Section~\ref{sec:ll-hl-equiv} shows a correspondence between the $m$-consumed property and a bound on the length of initialized chains, as well as a correspondence between the unseparated paths property and a bound on the length of marginalized chains.

\subsection{Low-Level Bounded Memory}
\label{sec:llstreaming}

A program executes in bounded memory if the delayed sampling graph maintains a bounded number of reachable variables over time.
We formalize this as follows:
\begin{definition}[Low-level Bounded-Memory]
    An execution $(s_n, g_n)_{n \in \mathbb{N}}$ of a model 
     is {\em low-level bounded-memory} if 
    $$
        \exists k. \; \forall n \geq 0 \; |\reachable(\dsgraph_n, s_n)| \le k * |\freerandvars(s_n)|
    $$
    \label{def:ll-streaming}
\end{definition}
This definition states that at each iteration, the size of the set of reachable nodes in the delayed sampling graph may be at most a constant multiple of the number of free random variables in the state.
We do not consider the runtime to violate bounded memory in the trivial case that the program state is intrinsically unbounded, i.e., when $|\freerandvars(s_n)|_{n\in\mathbb{N}}$ is unbounded.
Such a program would not execute in bounded memory under any inference algorithm; even a particle filter would require unbounded memory to store the program state.

\subsection{High-Level Definitions}
\label{sec:hlstreaming}

In this section, we present an alternative high-level definition of bounded memory that is easier to reason about.
The high-level definition is in terms of dataflow properties of delayed sampling operations.
We have formalized these dataflow properties by augmenting the delayed sampling operations with tracing.
A trace is defined as follows:
\begin{align*}
    \trace ::= & \; \trace :: \trace_1 \mid \texttt{nil}\\
    \trace_1 ::= & \; \dsnode \samplearrow \dsnode \mid \dsnode \samplearrow \texttt{nil} \mid \traceeval(\RVSet) \mid \traceobs (\dsnode)
\end{align*}

\pagebreak 
A trace is a list of primitive operations, where each primitive is one of:
\begin{itemize}
    \item Assumption, written $\dsnode \samplearrow \dsnode'$ when $\dsnode$ is assumed from another random variable $\dsnode'$ or $\dsnode \samplearrow \texttt{nil}$ when it is assumed without a parent.
    \item Evaluation using the $\traceeval$ keyword, which refers to evaluating a set of random variables $\RVSet$.
    \item Observation using the $\traceobs$ keyword, which refers to observing a random variable~$\dsnode$.
\end{itemize}
We define an augmented semantics that operates on a pair of a delayed sampling graph and a trace.
Figure~\ref{fig:augsem} defines augmented versions of the $\dsassume$, $\dsvalue$, and $\dsobserve$ operations, and the full semantics (written $\overline{\sem{\cdot}}$ and $\overline{\psem{\cdot}}$) is defined by replacing these operators in \Cref{fig:ds-sem} with their traced counterparts from \Cref{fig:augsem}.

\begin{figure*}
    $
    \begin{array}{lcl}
        \trassume(v, (g, \tau)) & = &  \letin{X', g' = \dsassume(v, g)} \\&&
\begin{cases}%
            X', (g', \tau :: X' \samplearrow \texttt{nil} ) & \freerandvars(v) = \emptyset \\%
            X', (g', \tau :: X' \samplearrow X ) & \{ X \} = \freerandvars(v) \wedge \conjugate(v, X, g)\\%
            X', (g', \tau :: \traceeval(\freerandvars(v)) :: X' \samplearrow \texttt{nil}) & \text{otherwise}%
        \end{cases}\\
        \trvalue(v, (g, \tau)) & = & \letin{(v', g') = \dsvalue(v, g)} v', (g', \tau :: \traceeval(\freerandvars(v))) \\
        \trobserve(X, v, (g, \tau)) & = & \dsobserve(X, v, g), (\tau :: \traceobs(X))
    \end{array}
    $
    \caption{Tracing semantics of delayed sampling operators.}
    \label{fig:augsem}
\end{figure*}

\paragraph{The $m$-consumed Property}
The \emph{$m$-consumed property} is used to enforce that every variable introduced with $\dsassume$ is eventually consumed either by directly being passed to a $\dsvalue$ or $\dsobserve$ or transitively by being passed to a $\dsassume$ that introduces a variable that is also $m$-consumed.
\begin{definition}[$m$-consumed]
    A variable $\dsnode$ is $m$-consumed in a trace $\trace$ under the following circumstances:
    \begin{itemize}
        \item $\dsnode$ is $0$-consumed if it is observed or evaluated (i.e., $\trace$ has $\traceeval(\dsnode)$ where $\dsnode \in \RVSet$ or $\traceobs(\dsnode)$).
        \item $\dsnode$ is $m$-consumed if it is passed to the $\dsassume$ statement that introduces another variable $\dsnode'$ (i.e., $\dsnode' \samplearrow \dsnode$ is in $\trace$), and $\dsnode'$ is $(m-1)$-consumed.
    \end{itemize}
\end{definition}

\paragraph{The Unseparated Paths Property}
The \emph{unseparated paths property} states the existence of a sequence of variables, each assumed from the previous, with no variable in the sequence observed or evaluated.
\begin{definition}[Unseparated Paths]
    An unseparated path in $\tau$ is a sequence of variables $\dsnode_0, \dsnode_1, \dots, \dsnode_n$ such that each $\dsnode_{i+1}$ was assumed from $\dsnode_i$ (i.e., $\dsnode_{i+1} \samplearrow \dsnode_i$ is in $\trace$) and no $\dsnode_i$ is directly observed or evaluated (i.e., $\trace$ does not contain any $\traceeval$ or $\traceobs$ operations that reference $\dsnode_i$).
\end{definition}

\paragraph{High-level Bounded Memory}
We now present the high-level bounded memory property.
This property states that all variables must eventually be $m$-consumed or unused, and there must be a uniform bound across iterations on the length of an unseparated path starting from a program state variable.

\begin{definition}[High-level Bounded-Memory]
    A program execution $(s_n, (\dsgraph_n, \trace_n))_{n \in \mathbb{N}}$ is \emph{high-level bounded-memory} if and only if
    \begin{itemize}
        \item There exists an $m$ such that for every iteration $n$ and every variable introduced before $n$ (i.e., $\dsnode$ such that $\dsnode \samplearrow \dsnode'$ or $\dsnode \samplearrow \texttt{nil}$ is in $\trace_n$), either a) there exists a $n' \ge n$ such that for all $n'' \ge n'$, $\dsnode$ is $m$-consumed in $\trace_{n''}$, or b) $\dsnode$ is unused -- meaning that for all $n' \ge n$, $X$ isn't an element of any unseparated path longer than $m$ in $\trace_{n'}$.
        \item There exists a $\notn$ such that for all $n$, no random variable referenced in $s_n$ starts an unseparated path in $\trace_n$ of length more than $\notn$.
    \end{itemize}
\end{definition}

\subsection{Equivalence of Low-Level and High-Level Definitions}
\label{sec:ll-hl-equiv}

In this section, we show the equivalence of the low-level and high-level definitions.
We do so by showing that both properties are equivalent to the delayed sampling graph having a uniform bound~(i.e., a bound that holds across all iterations) on the length of initialized and marginalized chains as defined in \Cref{subsec:ds_mem_usage}.

\subsubsection{Low-Level Bounded Memory vs. Infinite Chains}

\begin{lemma}
    \label{lem:llsound-helper}
    If the delayed sampling graph is constructed using $\dsassume$, $\dsobserve$, and $\dsvalue$ operations, then each random variable starts either
          an initialized chain,
          a marginalized chain, or
          an initialized chain followed by a marginalized chain.
\end{lemma}
\begin{proof}
  The  $\dsassume$, $\dsobserve$, and $\dsvalue$ operations can only make the following modifications to a delayed sampling graph~$g$.
  (1) Add a independent $\stmarginalized$ node which creates a marginalized chain of length zero.
  (2) Attach a new $\stinitialized$ node~$X$ to a node~$Y$ with a conjugate distribution. It means that $Y$ is either $\stinitialized$ or $\stmarginalized$ and thus it creates either a longer initialized chain or an initialized chain followed by a marginalized chain.
  (3) Perform a $\dsllgraft$ which ensures that every ancestor of a node is marginalized and has a single marginalized child. Every non-ancestor variable is either as it was before or becomes realized, so this operation preserves the structure of the previous graph and cannot increase the length of the chains.
  (4)~convert a $\stmarginalized$ node into a $\strealized$ node which can only break a chain.
\end{proof}

\begin{theorem}
    \label{thm:ll-chains}
    A program is low-level bounded-memory iff there is a uniform bound $m$ on the length of an initialized chain and a uniform bound $\notn$ on the length of a marginalized chain.
\end{theorem}
\begin{proof}
    Assuming a uniform bound, when the number of variables is bounded by $N$, according to Lemma~\ref{lem:llsound-helper}, the number of reachable nodes in the graph is bounded by $N \times (\notn + m)$.

    Conversely, if no uniform bound exists (i.e., for every potential bounds $\notn$ and $m$, there exists a iteration $n$ such that chains may exceed the bound at $n$), the execution cannot be low-level bounded-memory, because even if the number of root variables is bounded by $N$, the reachable variables may exceed $N \times (\notn + m)$.
\end{proof}

\subsubsection{High-Level Bounded Memory vs. Infinite Chains}

\begin{theorem}[High-level Soundness]
    \label{thm:hl-sound}
    In a program execution that is high-level bounded-memory, no infinite chains can exist in any of the delayed sampling graphs.
\end{theorem}

\begin{proof}
    All initialized chains must be shorter than $m$, where $m$ is from the $m$-consumed property of high-level bounded-memory.
    This is because when a variable's descendant is subject to $\dsobserve$ or $\dsvalue$, the variable becomes marginalized.
    Such a descendant can be at most $m$ variables away because of the definition of $m$-consumed.

    All marginalized chains must be shorter than $\notn + m$, where $\notn$ is from the unseparated path property of high-level bounded-memory and $m$ is from the $m$-consumed property.
    By \Cref{lem:llsound-helper}, every marginalized chain must start at either a root or an initialized chain.
    If it starts at a root, the unseparated path property ensures that the path between the root and the end of the chain can contain at most $\notn$ variables.
    This is because any observed or valued variables become realized and become the end of the chain.
    If it starts at an initialized chain, by the above reasoning that chain has length at most $m$, and there was a previous iteration at which the marginalized chain started at a root and had length at most $\notn$, giving an overall length of at most $\notn+m$.
\end{proof}

\begin{lemma}
    \label{lem:mpath-precise}
    If there exists a used variable that is not $m$-consumed, then the program produces a graph at some iteration with an initialized chain of length $m$.
\end{lemma}
\begin{proof}
    If a used variable is not $m$-consumed, then by the definition of $m$-consumed at some iteration it must start an assume chain of length $m$.
    All of the nodes in this chain must be initialized, and therefore form an initialized chain of length $m$.
\end{proof}

\begin{lemma}
    \label{lem:unseppath-precise}
    If every variable is $m$-consumed, and there exists a variable that starts an unseparated path of length $\notn$ where $\notn > m$, then there exists an iteration with a marginalized chain that has length at least $\notn - m$.
\end{lemma}
\begin{proof}
    Note that the first $\notn-m$ variables in the unseparated path must be either marginalized or realized. Otherwise, there would be more than $m$ initialized variables in the tail of the unseparated path that are initialized, which would violate soundness of $m$-consumed. Let $\dsnode$ be the variable that starts the unseparated path and $\dsnode'$ be the last marginalized or realized variable in the unseparated path, and consider the iteration $n'$ when $\dsnode'$ was first marginalized. It must be true that
          (1)~$\dsnode$ is in the program state at iteration $n'$ because it is in the state at the current iteration $n > n'$, and
          (2)~a marginalized chain runs from $\dsnode$ to $\dsnode'$.
    Thus, at $n'$, the marginalized chain had length $\notn-m$.
\end{proof}

\begin{theorem}[High-level Completeness]
    \label{thm:hl-precise}
    If a program execution is not high-level bounded-memory, the delayed sampling graph has either unbounded initialized chains or marginalized chains.
\end{theorem}
\begin{proof}
    If the execution is not high-level bounded-memory, it either fails the $m$-consumed property or the unseparated path property. If it fails the $m$-consumed property, apply Lemma~\ref{lem:mpath-precise}. Otherwise, apply Lemma~\ref{lem:unseppath-precise}.
\end{proof}

\begin{theorem}
    A program execution is high-level bounded-memory if and only if it is low-level bounded-memory.
    \label{thm:hl-ll-equiv}
\end{theorem}
\begin{proof}
    Apply Theorems \ref{thm:ll-chains}, \ref{thm:hl-sound}, and \ref{thm:hl-precise}.
\end{proof}

\section{Analysis}
\label{sec:analysis}

\newcommand{\graph}{\mathcal{G}}
\newcommand{\ctx}{\Gamma}
\newcommand{\hastypeabs}[6]{{#1}, {#2} \vdash_{#3} {#4} : {#5}, {#6}}
\newcommand{\hastype}[5]{\hastypeabs{#1}{#2}{}{#3}{#4}{#5}}
\newcommand{\hastypealpha}[5]{\hastypeabs{#1}{#2}{\alpha}{#3}{#4}{#5}}
\newcommand{\vhastypeabs}[4]{{#1} \vdash_{#2} {#3} : {#4}}
\newcommand{\vhastypealpha}[3]{\vhastypeabs{#1}{\alpha}{#2}{#3}}
\newcommand{\vhastypedet}[2]{\vhastypeabs{}{p}{#1}{#2}}
\newcommand{\sing}[1]{\mathcal{S}({#1})}
\newcommand{\pstream}[2]{\mathtt{stream}({#1}, {#2})}
\newcommand{\bounded}{\mathtt{bounded}}
\newcommand{\stepfn}[4]{\mathtt{stepfn}({#1},{#2},{#3},{#4})}

In this section, we develop an analysis to check that a \muF{} program executes in bounded memory.
We approach this problem by developing two independent analyses within a shared analysis framework.
One analysis checks the $m$-consumed property of a program and the other checks the unseparated paths property, which together ensure that the program executes in bounded memory (\Cref{sec:semantic-properties}).

Our shared analysis framework abstracts the execution of a program as the execution of abstract operations on an \emph{abstract graph}. 
An abstract graph abstracts the dynamic state of a program's delayed sampling graph.
We implement the analysis framework by means of a type system, such that well-typed programs satisfy the $m$-consumed and unseparated paths properties, given each analysis's respective instantiation of the abstract graph.
The typing judgment \[
    \hastypeabs{\ctx}{\graph}{\alpha}{e}{\type}{\graph'}
\] asserts that in a \emph{context} $\Gamma$, and for an abstract graph $\graph$, that an expression $e$ accesses the random variables denoted by the \emph{type} $t$ and yields a new abstract graph $\graph'$.   
The parameter $\alpha$ is either $\mathit{mc}$ to denote the $m$-consumed analysis or $\mathit{up}$ to denote the unseparated paths. We write $\vhastypealpha{\ctx}{e}{\type}$ as shorthand for $\hastypeabs{\ctx}{\graph}{\alpha}{e}{\type}{\graph}$ when $e$ has no effect on the graph.

\subsection{Types and Contexts}

A \emph{type} $t$ captures the random variables the expression could refer to as well as its shape, as primitive data, a product, a function, or a stream instance.
\begin{align*}
    \type &\Coloneqq r \mid () \mid \type_1 \times \type_2 \mid t_1 \to t_2 \mid \pstream{\type}{s} \mid \bounded \\
    s &\Coloneqq \stepfn{p_{\mathit{state}}}{p_{\mathit{in}}}{\ctx_e}{e}
\end{align*}
The type of a primitive expression is a \emph{reference set}, denoted $r$, which specifies the random variables to which the expression refers.
We distinguish two types of stream instances, before and after bounded-memory checking. The first is $\pstream{\type}{s}$, where $\type$ is the type of the current state and $s$ is a step function representation to be described later. The second is $\bounded$, representing instances that have passed bounded memory analysis and hide their inner structure.

\paragraph{Reference Sets}

A \emph{reference set} of a \muF{} expression, denoted $r$, specifies the random variables that are affected when the expression is observed or evaluated. In the presence of branches, we define $r$ to be a pair of sets $(\mathit{lb}, \mathit{ub})$, where the lower bound $\mathit{lb}$ contains all random variables which \emph{must} be affected and the upper bound $\mathit{ub}$ all random variables which \emph{may} be affected. 
For example, a constant value in \muF{} such as $1.5$ has the reference set $(\emptyset, \emptyset)$ because it references no random variables. If the program variables $\mathtt{x}$ and $\mathtt{y}$ correspond to random variables $X$ and $Y$ respectively, then the expression \zl{gaussian(x,y)}, specifying a distribution with two parameters, has reference set $(\{X, Y\}, \{X, Y\})$, meaning that observing it will observe the random variables $X$ and $Y$.

\paragraph{Contexts}

The context $\ctx, x : \type$ maps variable $x$ to type $\type$. As \muF{} syntactic patterns $p$ may be variables or pairs, we use the shorthand $\ctx, p : \type$ to define types for variables in $p$ by structural correspondence with $\type$, as defined by the first rule below. We also define a judgment $\vdash_p p : \type$ that synthesizes a deterministic type $t$ from a pattern $p$.
\begin{mathpar}
    \inferrule{\ctx, p_1 : \type_1, p_2 : \type_2 \vdash_\alpha e : \type}{\ctx, (p_1, p_2) : \type_1 \times \type_2 \vdash_\alpha e : \type}

    \inferrule{ }{\vhastypedet{x}{(\emptyset, \emptyset)}}

    \inferrule{\vhastypedet{p_1}{\type_1} \and \vhastypedet{p_2}{\type_2}}{\vhastypedet{\muFpair{p_1}{p_2}}{\type_1 \times \type_2}}
\end{mathpar}

\subsection{Abstract Graphs}
\label{subsec:absgraph}

\newcommand{\assumeabs}[1]{{\mathit{assume}}_{#1}}
\newcommand{\observeabs}[1]{{\mathit{observe}}_{#1}}
\newcommand{\valueabs}[1]{{\mathit{value}}_{#1}}
\newcommand{\assumej}[4]{\assumeabs{#1}({#2}, {#3}, {#4})}
\newcommand{\observej}[4]{\observeabs{#1}({#2}, {#3}, {#4})}
\newcommand{\valuej}[3]{\valueabs{#1}({#2}, {#3})}

An \emph{abstract graph} $\graph$ is an abstraction of the delayed sampling graph that tracks which random variables have been consumed and active paths between random variables, properties relevant to the semantic properties.
For each analysis $\alpha$ there exists an abstract graph type, $\graph$, and a set of operations that form its interface (\Cref{fig:graph-interface}).

Specifically, in the $m$-consumed analysis we define $\graph$ to be a pair of sets $\mathit{in}$ and $\mathit{con}$ which respectively represent an over-approximation of variables introduced into the graph and an under-approximation of the variables consumed by observation or sampling (\Cref{fig:mc-graph}).
In the unseparated paths analysis, we define $\graph$ to be a set $\mathit{sep}$ of separators containing consumed random variables and a partial path function $p$ mapping a pair of random variables to an upper bound on the length of an unseparated path between them (\Cref{fig:up-graph}).

Operations on the abstract graph manipulate random variables, graphs, and reference sets.
The function $\dsassume$ returns a new graph with a random variable $X$ from a distribution with reference set $r$ added to $\graph$, $\dsobserve$ returns a graph where $X$ is observed with a value with reference set $r$, and $\dsvalue$ returns a graph where an expression with reference set $r$ is evaluated. The join operator $\sqcup_{\alpha}$ represents a conservative choice between two graphs.

\paragraph{$m$-consumed Graph Operations}

\begin{figure}
\centering
\begin{minipage}[b][][b]{0.4\textwidth}
\centering
\begin{align*}
    \assumeabs{\alpha} &: \mathrm{RV} \to r \to \graph \to \graph \\
    \observeabs{\alpha} &: \mathrm{RV} \to r \to \graph \to \graph \\
    \valueabs{\alpha} &: r \to \graph \to \graph \\
    \sqcup_\alpha &: \graph \to \graph \to \graph
\end{align*}
\caption{Abstract graph interface.} \label{fig:graph-interface}
\end{minipage}\hfill
\begin{minipage}[b][][b]{0.6\textwidth}
\centering
\begin{align*}
    \graph \Coloneqq{}& \{in \subseteq \textrm{RV}; \mathit{con} \subseteq \textrm{RV}\} \\
    \assumej{\mathit{mc}}{X}{r}{\graph} ={}& \{\graph.in \cup \{X\}; \graph.\mathit{con} \cup r.\mathit{lb}\} \\
    \observej{\mathit{mc}}{X}{r}{\graph} ={}& \{\graph.in; \graph.\mathit{con} \cup r.\mathit{lb} \cup \{X\}\} \\
    \valuej{\mathit{mc}}{r}{\graph} ={}& \{\graph.in; \graph.\mathit{con} \cup r.\mathit{lb}\} \\
    \graph_1 \sqcup_{\mathit{mc}} \graph_2 ={}& \{\graph_1.in \cup \graph_2.in; \graph_1.\mathit{con} \cap \graph_2.\mathit{con}\}
\end{align*}
\caption{$m$-consumed abstract graph operations.} \label{fig:mc-graph}
\end{minipage}
\end{figure}
In \Cref{fig:mc-graph}, $\assumej{\mathit{mc}}{X}{r}{\graph}$ marks the random variable~$X$ as introduced.
In all cases, the lower bound of random variables in the input is marked consumed. To join two states, we union the introduced variables and intersect the consumed variables.

\paragraph{Unseparated Paths Graph Operations}

\begin{figure}
\begin{align*}
    \graph \Coloneqq{} & \{p : \mathrm{RV} \times \mathrm{RV} \hookrightarrow \mathbb{N}; \mathit{sep} \subseteq \textrm{RV}\} \\
    \assumej{\mathit{up}}{X}{r}{\graph} ={} & \{p'; \graph.\mathit{sep}\} \textrm{ where } p'(X, X) \mapsto 0, \\
    & \quad p'(X_i, X) \mapsto \graph.p(X_i, X_p) + 1 \textrm{ for all } X_p \in r.\mathit{ub} \setminus \graph.\mathit{sep}, X_i \in \textrm{RV}, \\
    & \quad p'(X, Y) \mapsto \graph.p(X, Y)\ \textrm{otherwise} \\
    \observej{\mathit{up}}{X}{r}{\graph} ={} & \{\graph.p; \graph.\mathit{sep} \cup r.\mathit{lb} \cup \{X\}\} \\
    \valuej{\mathit{up}}{r}{\graph} ={} & \{\graph.p; \graph.\mathit{sep} \cup r.\mathit{lb}\} \\
    \graph_1 \sqcup_{\mathit{up}} \graph_2 ={} & \{p'; \graph_1.\mathit{sep} \cap \graph_2.\mathit{sep}\} \textrm{ where } p'(v_1, v_2) \mapsto \max(\graph_1.p(v_1, v_2), \graph_2.p(v_1, v_2))
\end{align*}
\caption{Unseparated paths abstract graph operations.} \label{fig:up-graph}
\end{figure}
In \Cref{fig:up-graph}, $\observeabs{\mathit{up}}$ and $\valueabs{\mathit{up}}$ mark input variables as separators. In $\assumeabs{\mathit{up}}$, we set the length of the path from the new variable $X$ to itself to zero. For a parent $X_p$ that is not a separator, we set the length of the path from any variable $X_i$ to $X$ to one more than the length from $X_i$ to $X_p$. To join two states, we intersect the separators and take the maximum length between the results of the two path functions (where defined, or 0 otherwise). 

\subsection{Typing Rules}
\label{subsec:types}

\newcommand{\freshvar}[2]{{#1} = \textsf{fresh}({#2})}
\newcommand{\muFlam}[2]{\ensuremath{\texttt{fun }#1}\texttt{ -> }#2}

\newcommand{\mvalid}[4]{\ensuremath{{#1}, {#2} \vdash_{#3} \texttt{infer}({#4})}}
\newcommand{\rfold}{\searrow}

\newcommand{\erasure}[1]{\ensuremath{\overline{#1}}}

\begin{figure}
    \small
    \begin{mathpar}
        \inferrule{ }{\vhastypealpha{\ctx}{c}{(\emptyset,\emptyset)}}

        \inferrule{\vhastypealpha{\ctx}{v}{r} \and \freshvar{X}{\graph}}{\hastypealpha{\ctx}{\graph}{\muFsample{v}}{(\{X\},\{X\})}{\assumej{\alpha}{X}{r}{\graph}}}

        \inferrule{\hastypealpha{\ctx}{\graph}{\muFsample{v_1}}{(\{X\},\{X\})}{\graph'} \and \vhastypealpha{\ctx}{v_2}{r_2}}{\hastypealpha{\ctx}{\graph}{\muFobserve{v_1}{v_2}}{()}{\observej{\alpha}{X}{r_2}{\valuej{\alpha}{r_2}{\graph'}}}}

        \inferrule{\vhastypealpha{\ctx}{v}{\type} \and \type \rfold r}{\vhastypealpha{\ctx}{\muFop{v}}{r}}

        \inferrule{\hastypealpha{\ctx}{\graph}{e}{\type}{\graph'} \and \hastypealpha{\ctx, p : \type}{\graph'}{e'}{\type'}{\graph''}}{\hastypealpha{\ctx}{\graph}{\muFletin{p}{e}{e'}}{\type'}{\graph''}}

        \inferrule{\vhastypealpha{\ctx}{v}{r} \and \graph' = \valuej{\alpha}{r}{\graph} \and \hastypealpha{\ctx}{\graph'}{e_1}{\type_1}{\graph_1} \and \hastypealpha{\ctx}{\graph'}{e_2}{\type_2}{\graph_2}}{\hastypealpha{\ctx}{\graph}{\muFif{v}{e_1}{e_2}}{\type_1 \sqcup \type_2}{\graph_1 \sqcup_\alpha \graph_2}}

        \inferrule{\vhastypealpha{\ctx}{m}{(\type, s)}}{\vhastypealpha{\ctx}{\muFinit{m}}{\pstream{\type}{s}}}

        \inferrule{\vhastypealpha{\ctx}{x}{\pstream{\type}{\stepfn{p_{\mathit{state}}}{p_{\mathit{in}}}{\ctx_e}{e}}} \\ \vhastypealpha{\ctx}{v}{\type_{\mathit{in}}} \\ \hastypealpha{\ctx_e, p_{\mathit{state}} : \type, p_{\mathit{in}} : \type_{\mathit{in}}}{\graph}{e}{\type' \times \type_{\mathit{out}}}{\graph'}}{\hastypealpha{\ctx}{\graph}{\muFunfold{x}{v}}{\type_{\mathit{out}} \times \pstream{\type'}{\stepfn{p_{\mathit{state}}}{p_{\mathit{in}}}{\ctx_e}{e}}}{\graph'}}

        \inferrule{\ctx \vdash_{\mathit{mc}} m\ \bounded \\ \ctx \vdash_{\mathit{up}} m\ \bounded}{\vhastypealpha{\ctx}{\muFinfer{m}}{\bounded}}

        \inferrule{\vhastypealpha{\ctx}{x}{\bounded} \and \vhastypealpha{\ctx}{v}{\type} \and \type \rfold (\emptyset, \emptyset)}{\vhastypealpha{\ctx}{\muFunfold{x}{v}}{(\emptyset,\emptyset) \times \bounded}}
    \end{mathpar}
    \normalsize
\caption{Delayed sampling type system.}
\label{fig:ds-types}
\end{figure}

In \Cref{fig:ds-types} we present the typing rules that are relevant to analyzing probabilistic streams, with the full definition in \Cref{sec:types-app}.
Constants reference no random variables. $\zl{sample}$ introduces a fresh random variable sampled from its argument and adds it to the graph. $\zl{observe}$ introduces an intermediate random variable for its first argument by the same mechanism as $\zl{sample}$, and observes it to be the evaluation of its second argument. 

\paragraph{Operators and Scalar Folding} 

We use \muF{} operators $op$ to describe probability distributions and other operations over scalars and assume them to have scalar return values. The auxiliary judgment $\rfold$ folds products and stream instances into scalars by taking unions of variable sets.

\footnotesize
\begin{mathpar}
    \inferrule{ }{() \rfold (\emptyset, \emptyset)}

    \inferrule{ }{r \rfold r}

    \inferrule{\type_1 \rfold (\mathit{lb}, \mathit{ub}) \quad \type_2 \rfold (\mathit{lb}', \mathit{ub}')}{\type_1 \times \type_2 \rfold (\mathit{lb} \cup \mathit{lb}', \mathit{ub} \cup \mathit{ub}')}

    \inferrule{\type \rfold (\mathit{lb}, \mathit{ub})}{\pstream{\type}{s} \rfold (\mathit{lb}, \mathit{ub})}

    \inferrule{ }{\bounded \rfold (\emptyset, \emptyset)}
\end{mathpar}
\normalsize

\paragraph{Sequencing}

Sequencing using the $\zl{let}$-expression follows the standard typing rule for $\zl{let}$, and also threads the output graph of evaluating $e$ into the evaluation of $e'$.

\paragraph{Conditionals and Join}

\newcommand{\measuredparboxalign}[1]{\parbox{\widthof{$\!\begin{aligned}#1\end{aligned}$}}{$\!\begin{aligned}#1\end{aligned}$}}

\begin{wrapfigure}{l}{0.45\textwidth}
    \centering
    \vspace{-10pt}
    \measuredparboxalign{
        () \sqcup () &= () \\
        (\type_1 \times \type_2) \sqcup (\type'_1 \times \type'_2) &= (\type_1 \sqcup \type'_1) \times (\type_2 \sqcup \type'_2) \\
        (\mathit{lb}, \mathit{ub}) \sqcup (\mathit{lb}', \mathit{ub}') &= (\mathit{lb} \cap \mathit{lb}', \mathit{ub} \cup \mathit{ub}')
    }
    \caption{Join operator for types.}
    \label{fig:join}
\end{wrapfigure}
\zl{if}-expressions evaluate the condition, check both branches in parallel, and join the resulting reference set and graphs.
The join operator $\sqcup$ (\Cref{fig:join}), representing the conservative union of two types, unions the upper bounds and intersects the lower bounds. We disallow \zl{if}-branching over functions and stream instances. 

\pagebreak 

\paragraph{Streams and Inference}

To facilitate typing of stream functions, we define the following auxiliary judgment, which computes, for a stream function, the type of its initial state and the syntactic fragment for its step function.
\begin{mathpar}
    \inferrule{\vhastypealpha{\ctx}{e'}{\type_{\mathit{init}}} \\ \type_{\mathit{init}} \rfold (\emptyset, \emptyset)}{\vhastypealpha{\ctx}{\muFstream{e'}{p_{\mathit{state}}}{p_{\mathit{in}}}{e}}{(\type_{\mathit{init}}, \stepfn{p_{\mathit{state}}}{p_{\mathit{in}}}{\ctx}{e})}}
\end{mathpar}
Correspondingly, we define the context $\ctx, m : (\type_{\mathit{init}}, \stepfn{p_{\mathit{state}}}{p_{\mathit{in}}}{\ctx_e}{e})$ to map the stream function name $m$ to its initial state type and step function.

Instances that are created by \zl{init} expose the type of their internal state and their step function.
The \zl{unfold} rule applies the step function to the current state, yielding an output and an instance with the new state. It ensures that the argument $v$ is compatible with the type of the step function.

An $\zl{infer}$ expression marks the entry point of a new sub-analysis for its new delayed sampling graph. The premises of the typing rule for \zl{infer} are the \emph{success conditions} for both analyses that must hold regardless of $\alpha$. This judgment, $\ctx \vdash_{\alpha} m\ \bounded$, states that the stream function $m$ can be unfolded for an arbitrary number of iterations while satisfying property $\alpha$ starting with an empty delayed sampling graph. 

Instances created by \zl{infer} possess a newly instantiated delayed sampling graph. Their internal state contains the delayed sampling graph and bookkeeping information for the inference algorithm. Thus, the state is hidden to the exterior and the instance is assigned the opaque type $\bounded$. \zl{unfold} on a $\bounded$ type only requires that the input and output are purely deterministic.

\paragraph{\emph{m}-consumed Success Condition}

We conclude a stream function passes the $m$-consumed analysis when all variables that are introduced are consumed by the program. Because an introduced variable may take several stream iterations to be consumed, we repeatedly execute the analysis until we consume all variables and succeed or reach a fixed point and fail. Define the iteration judgment $\ctx \vdash_{\alpha(n)} m : \type', \graph$, where $\alpha$ is either $mc$ or $up$, as follows:
\begin{mathpar}
    \inferrule{\vhastypeabs{\ctx}{\alpha}{m}{(\type, \stepfn{p_{\mathit{state}}}{p_{\mathit{in}}}{\ctx_e}{e})} \\ \vhastypedet{p_{\mathit{in}}}{\type_{\mathit{in}}} \and \hastypeabs{\ctx_e, p_{\mathit{in}} : \type_{\mathit{in}}, p_{\mathit{state}} : \type}{\bot_\alpha}{\alpha}{e}{\type_{\mathit{out}} \times \type'}{\graph}}{\ctx \vdash_{\alpha(0)} m : \type', \graph}

    \inferrule{\vhastypeabs{\ctx}{\alpha}{m}{(\type, \stepfn{p_{\mathit{state}}}{p_{\mathit{in}}}{\ctx_e}{e})} \and \ctx \vdash_{\alpha(n-1)} m : \type', \graph \\ \vhastypedet{p_{\mathit{in}}}{\type_{\mathit{in}}} \and \hastypeabs{\ctx_e, p_{\mathit{in}} : \type_{\mathit{in}}, p_{\mathit{state}} : \type'}{\graph}{\alpha}{e}{\type_{\mathit{out}} \times \type''}{\graph'}}{\ctx \vdash_{\alpha(n)} m : \type'', \graph'}
\end{mathpar}
On each iteration, this judgment applies the appropriate type rule for the step function and returns the result, using the abstract graph from the previous iteration as the context for the step function rule. The initial iteration uses an empty abstract graph as the context, represented by $\bot_\alpha$. For the $m$-consumed analysis, we specialize the judgment to $\ctx \vdash_{mc(n)} m : \type', \graph$, and define $\bot_{\mathit{mc}}$ to be $(\emptyset, \emptyset)$.

The rule continues iterating until it reaches the success condition. The success condition states that every variable introduced that is kept in the program state must be used with in a bounded number of time steps. We formalize this as the following type rule:
\begin{mathpar}
    \inferrule{\ctx \vdash_{mc(0)} m : \type, \graph \and t \rfold (lb, ub) \and \ctx \vdash_{mc(n)} m : \type'', \graph' \and (\graph.in \setminus \graph'.\mathit{con}) \cap \mathit{ub} = \emptyset}{\ctx \vdash_{\mathit{mc}} m\ \bounded}
\end{mathpar}
Alternatively, if evaluating one more iteration does not consume any more variables, we reach a fixed point and return failure. Since every iteration we either consume a variable or reach a fixed point, the analysis is guaranteed to terminate.

\paragraph{Unseparated Paths Success Condition}

\newcommand{\size}{\textrm{size}}

Like the $m$-consumed analysis, the unseparated paths analysis is iterative, and we may need to repeat it for some number of iterations. 
We specialize the iteration judgment defined in the previous section to $\ctx \vdash_{up(n)} m : \type, \graph$ and define $\bot_{\mathit{up}}$ to be a pair of an empty map and an empty set. Define $\textrm{path}(\type, \graph)$ where $\type \rfold (\mathit{lb}, \mathit{ub})$ to be the length of the longest path from any random variable in $\mathit{ub}$ to any other variable in $\graph.p$. Then we conclude the program passes the unseparated path analysis when the length of the longest path converges after some finite number of iterations:
\begin{mathpar}
    \inferrule{\ctx \vdash_{up(n)} m : \type, \graph \quad \ctx \vdash_{up(n + (\textrm{path}(\type, \graph) * \size(\type)) + 1)} m : \type'', \graph' \quad \textrm{path}(\type, \graph) = \textrm{path}(\type'', \graph')}{\ctx \vdash_{\mathit{up}} m\ \bounded}
\end{mathpar}
The implementation of this rule repeatedly computes a new abstract graph starting from the previous iteration's output. 
It exits when the longest path length at the current iteration is equal to the longest path after $(\textrm{path}(\type, \graph) * \size(\type)) + 1$ additional iterations.
The function $\size$ determines, for a given type $t$, how many values of base type are contained in $t$.
\begin{mathpar}
    \size(r) = 1 \and 
    \size(\type_1 \times \type_2) = \size(\type_1) + \size(\type_2) 
\end{mathpar}
The extra iterations ensure that the path length has stabilized and the analysis can safely conclude that there is a bound on the length of the longest unseparated path.

If the path length check fails, the implementation keeps iterating until a pre-specified bound is reached. Upon reaching this bound, the implementation outputs an analysis failure.
Note that the analysis may be imprecise and reject correct programs if the bound is not sufficiently high.

\subsection{Example Type Derivation}

This section presents an example type derivation used by the analysis to confirm that the program in~\Cref{fig:lqr_src} satisfies the $m$-consumed property. In particular, we confirm that the stream function \texttt{kalman} passes the $m$-consumed analysis.

Using $e$ as shorthand for the body of its step function, we derive the following $m$-consumed success condition for \texttt{kalman}:
\begin{mathpar}
\inferrule{
    \vhastypeabs{\ctx}{\mathit{mc}(0)}{\texttt{kalman}}{((\{X\}, \{X\}) \times (\{X\}, \{X\}))} \\
    ((\{X\}, \{X\}) \times (\{X\}, \{X\})) \rfold (\{X\}, \{X\}) \\
    \vhastypeabs{\ctx}{\mathit{mc}(0)}{\texttt{kalman}}{((\{X\}, \{X\}) \times (\{X\}, \{X\}))} \\
    (\{X\} \setminus \{X\}) \cap \{X\} = \emptyset
}
{\ctx \vdash_{\mathit{mc}} \texttt{kalman}\ \bounded}
\end{mathpar}
The second and fourth premises follow immediately from definitions and set operations.
The derivation of the first and third premises is as follows:
\begin{mathpar}
\inferrule{
    \vhastypeabs{\ctx}{\mathit{mc}}{\texttt{kalman}}{((\emptyset, \emptyset), \stepfn{\texttt{pre\_x}}{\texttt{obs}}{\Gamma_e}{e})} \\
    \vhastypedet{\texttt{obs}}{(\emptyset, \emptyset)} \\\\
    \hastypeabs{\Gamma_e, \texttt{obs} : (\emptyset, \emptyset), \texttt{pre\_x} : (\emptyset, \emptyset)}{(\emptyset,\emptyset)}{\mathit{mc}}{e}{((\{X\}, \{X\}) \times (\{X\}, \{X\}))}{(\{X\}, \{X\})}\\
}
{\vhastypeabs{\ctx}{\mathit{mc}(0)}{\texttt{kalman}}{((\{X\}, \{X\}) \times (\{X\}, \{X\}))}}
\end{mathpar}
The second premise follows from definitions. The derivation of the first premise is as follows:
\begin{mathpar}
\inferrule{
    \vhastypeabs{\Gamma_e}{\mathit{mc}}{0.0}{(\emptyset, \emptyset)} \\
    (\emptyset, \emptyset) \rfold (\emptyset, \emptyset)
}{\vhastypeabs{\ctx}{\mathit{mc}}{\muFstream{0.0}{\texttt{pre\_x}}{\texttt{obs}}{e}}{((\emptyset, \emptyset), \stepfn{\texttt{pre\_x}}{\texttt{obs}}{\Gamma_e}{e})}}
\end{mathpar}
where the premises follow immediately. Finally, let $\ctx'$ be the context $\Gamma_e, \texttt{obs} : (\emptyset, \emptyset), \texttt{pre\_x} : (\emptyset, \emptyset)$ and $t$ be the type $(\emptyset, \emptyset)$. The derivation of the third premise is as follows.
\begin{mathpar}
\inferrule{
    \hastypeabs{\ctx'}{(\emptyset, \emptyset)}{mc}{\muFsample{\texttt{gaussian}(\texttt{pre\_x}, 1.0)}}{t}{(\{X\}, \emptyset)} \\
    \hastypeabs{\ctx', \texttt{x} : t}{(\{X\}, \emptyset)}{mc}{\muFletin{()}{\muFobserve{\texttt{gaussian}(\texttt{x}, 1.0)}{\texttt{obs}}}{(\texttt{x}, \texttt{x})}}{(t \times t)}{(\{X\}, \{X\})}
}{\hastypeabs{\ctx'}{(\emptyset, \emptyset)}{mc}{e}{(t \times t)}{(\{X\}, \{X\})}}
\end{mathpar}
The first premise follows from the typing rule for \texttt{sample}. The second premise follows from the typing rule for \texttt{let} as follows:
\begin{mathpar}
\inferrule{
    \hastypeabs{\ctx', \texttt{x} : t}{(\{X\}, \emptyset)}{mc}{\muFobserve{\texttt{gaussian}(\texttt{x}, 1.0)}{\texttt{obs}}}{()}{(\{X\}, \{X\})} \\
    \hastypeabs{\ctx', \texttt{x} : t}{(\{X\}, \{X\})}{mc}{(\texttt{x}, \texttt{x})}{(t \times t)}{(\{X\}, \{X\})}
}{\hastypeabs{\ctx', \texttt{x} : t}{(\{X\}, \emptyset)}{mc}{\muFletin{()}{\muFobserve{\texttt{gaussian}(\texttt{x}, 1.0)}{\texttt{obs}}}{(\texttt{x}, \texttt{x})}}{(t \times t)}{(\{X\}, \{X\})}}
\end{mathpar}
where the first premise follows from the rule for \texttt{observe} and the second from the rule for pairs.

\subsection{Soundness}
\label{subsec:soundness}

Here, we outline how we show the type system is sound.
We give a high-level overview of the approach; the details are in \Cref{app:soundness}.

\paragraph{Entailment Relations}
In \Cref{subsec:typesapp}, we establish several {\em entailment relations} that relate semantic objects to their type-level counterparts.
These relations are parameterized by $\alpha$ which is either $mc$ for the $m$-consumed relation or $up$ for the unseparated path relation.
We write $v \vDash_\alpha t$ to mean a value entails a type.
We write $\gamma \vDash_\alpha \Gamma$ to mean an environment entails a type context.
We write $v, (g, \trace) \vDash_\alpha \type, \graph$ to mean a value and traced graph (see \Cref{sec:hlstreaming} for the definition of a traced graph) entail a type and abstract graph.
We write $\gamma, (g, \trace) \vDash_\alpha \Gamma, \graph$ to mean an environment and traced graph entail a type context and abstract graph.

\paragraph{Soundness}
The following theorems establish the soundness of the type system. 
The first theorem states that the type system soundly ascribes types to values and soundly updates the abstract delayed sampling graph:
\begin{theorem}[$m$-consumed and Unseparated Path Soundness]
    If $\gamma, (g, \trace) \vDash_\alpha \Gamma, \graph$ and 
       $\Gamma, \graph \vdash_\alpha e : t, \graph'$ and 
       $\overline{\psem{e}}_\gamma(g, \trace), w = v, (g', \trace'), w'$, then 
       $v, (g', \trace') \vDash_\alpha t, \graph'$.
    \label{thm:mc-up-sound}
\end{theorem}
\noindent Next, the type system soundly ensures a stream function maintains bounded memory.
\begin{theorem}[Analysis Soundness]
    \label{thm:ana-sound}
    If $\gamma \vDash_\alpha \Gamma$ and $\Gamma \vdash_\alpha m : \bounded$, then $\overline{\sem{m}}_\gamma \vDash_\alpha \bounded$.
\end{theorem}
\noindent We prove these theorems in \Cref{subsec:typesapp}.

\subsection{Implementation}

We implemented our analysis framework and the $m$-consumed and unseparated paths analyses in OCaml. Our implementation takes as input a \muF{} program and outputs either true or false for each analysis. It also accepts a parameter for the iteration bound for the unseparated paths analysis. The implementation goes beyond the type system laid out in the paper by supporting functions that have probabilistic effects as well as interfaces for list and array operations.
\muF{} programs can further be compiled to OCaml and executed using the ProbZelus delayed sampling runtime.
The code is available at \url{https://github.com/psg-mit/probzelus-oopsla21}.

\section{Evaluation}
\label{sec:evaluation}

To evaluate the ability of the analysis to accept only \muF{} programs that can execute in bounded memory, we executed it on several benchmarks reflective of real-world inference tasks. 

\emph{Research Questions}. We used our implementation to answer two research questions. For realistic probabilistic programs, (1) does the type system precisely verify the  properties required for bounded-memory execution, and (2) is a small iteration bound sufficient for the unseparated paths analysis?

\subsection{Methodology}

We executed the analysis on example programs from \citet{probzelus} originally written in \ProbZelus{}, a probabilistic programming language featuring probabilistic data streams and delayed sampling. We manually translated them to \muF{}, and they reflect a range of realistic control problems with different memory usage characteristics. For the unseparated paths analysis, we set an iteration count bound of 10, which was sufficient for these programs. We compared the outputs of the analysis to our manual logical reasoning about the ability of each of the following programs to execute in bounded memory. We provide source code for all benchmarks in \Cref{sec:benchmarks}.

\emph{Kalman} is the simplified core model of \Cref{fig:lqr_src} and models an agent that estimates position from noisy observations.
Applying delayed sampling on this model is equivalent to a Kalman filter \cite{kalman1960} where each particle returns the exact solution. 

\emph{Kalman Hold-First} is the example from \Cref{fig:kalman_first_src} with a reference to the output of the first iteration.

\emph{Gaussian Random Walk} is a simplification of Kalman that does not observe of the true position, effectively expressing a Gaussian random walk.

\emph{Robot} is the full example from \Cref{fig:lqr_src} that includes the Kalman core model as well as a main stream function that invokes a controller based on the inferred position.

\emph{Coin} models an agent that estimates the bias of a coin. The model chooses the probability of the coin from a uniform distribution, and thereafter chooses the observations by flipping a coin with that probability. Applying delayed sampling to this model is equivalent to exact inference in a Beta-Bernoulli conjugate model \cite{fink1997} where each particle returns the exact solution.

\emph{Gaussian-Gaussian} estimates the mean and variance of a Gaussian distribution.

\emph{Outlier}, adapted from Section 2 of \cite{minka2001}, models the same situation as the Kalman benchmark, but with a sensor that can occasionally produce invalid readings. The model chooses the probability of an invalid reading from a $\zl{beta(100,1000)}$ distribution, so that invalid readings occur approximately 10\% of the time. At each time step, with the previously chosen probability, the model chooses the observation from either the invalid distribution $\zl{gaussian(0,100)}$ or the Kalman model. Applying delayed sampling to this model is equivalent to a Rao-Blackwellized particle filter \cite{doucet2000} combining exact inference with approximate particle filtering.

\emph{MTT (Multi-Target Tracker)} is adapted from \cite{mtt} and involves a variable number of targets with linear-Gaussian 2D position/velocity motion models that produce measurements of position at each time step. The model randomly introduces targets as a Poisson process and deletes them with fixed probability at each step.

\emph{SLAM (Simultaneous Localization and Mapping)} is adapted from \cite{slam} and models an agent that estimates its position on a one-dimensional grid and also a map of its environment associating each cell with black or white. The robot uses inference to decide its next move, but its motion commands are noisy with some probability that its wheels may slip, and its observations may also be incorrectly reported.

\subsection{Analysis Results}

\begin{table}[t]
\caption{Bounded memory analysis on benchmark programs.}
\begin{tabular}{ l c c c c c c }
\toprule
& \multicolumn{2}{c}{$m$-consumed} & \multicolumn{2}{c}{unsep. paths} & \multicolumn{2}{c}{bounded mem.} \\
\cmidrule{2-7}
& output & actual & output & actual & output & actual \\
\midrule
Kalman & \checkmark & \checkmark & \checkmark & \checkmark & \checkmark & \checkmark \\
Kalman Hold-First & \checkmark & \checkmark & \xmark & \xmark & \xmark & \xmark \\
Gaussian Random Walk & \xmark & \xmark & \checkmark & \checkmark & \xmark & \xmark \\
Robot & \checkmark & \checkmark & \checkmark & \checkmark & \checkmark & \checkmark \\
Coin & \checkmark & \checkmark & \checkmark & \checkmark & \checkmark & \checkmark \\
Gaussian-Gaussian & \checkmark & \checkmark & \checkmark & \checkmark & \checkmark & \checkmark \\
Outlier & \xmark & \xmark & \checkmark & \checkmark & \xmark & \xmark \\
MTT & \xmark & \xmark & \checkmark & \checkmark & \xmark & \xmark \\
SLAM & \xmark & \checkmark & \checkmark & \checkmark & \xmark & \checkmark \\
\bottomrule
\end{tabular}
\label{tbl:results}
\end{table}

Table~\ref{tbl:results} displays the analysis outputs for each of the benchmark programs.
For each analysis, the ``output'' column is the result of the implementation, and the ``actual'' column is the ground truth, i.e., whether the program satisfies the semantic property according to manual analysis. The ``bounded memory'' columns are the logical conjunction of the two semantic properties.

For the first six benchmarks, the analysis implementation yielded the same answer as manual analysis for whether the program satisfies both semantic properties and thus permits execution in bounded memory. In every case, the output of the implementation is sound with respect to the ground truth. Furthermore, all unseparated-path analyses converged within 10 iterations.

\emph{Kalman}. For this program, every variable is $m$-consumed for $m \le 1$ and starts an unseparated path of length at most 1, and thus it can execute in bounded memory.

\emph{Kalman Hold-First}. For this program, every variable is $m$-consumed for $m \le 1$. However, the analysis detects that unseparated paths starting from the initial value for $\zl{x}$ grow without bound and fail to converge after 10 iterations, so this program cannot execute in bounded memory.

\emph{Gaussian Random Walk}. Here, every unseparated path has length at most 1. However, the analysis detects that there is no $m$ such that any variable is $m$-consumed because no variable is ever observed or evaluated, so this program cannot execute in bounded memory.

\emph{Robot}. Every variable is $1$-consumed and every separated path has length at most 1. The analysis succeeds and indicates this program can execute in bounded memory.

\emph{Coin}. Every variable is $1$-consumed and every unseparated path has length at most 1. The analysis succeeds and indicates this program can execute in bounded memory.

\emph{Gaussian-Gaussian}. Every variable is $1$-consumed and every separated path has length at most 1. The analysis succeeds and indicates this program can execute in bounded memory.

\emph{Outlier}. Every unseparated path has length at most 1. However, in the event that samples are indefinitely considered outliers, no observation will occur that causes the variable $\zl{xt}$ to be consumed, so this program cannot execute in bounded memory.

\emph{MTT}. Every unseparated path has length at most 1. However, not all random variables are guaranteed to be consumed, as the final \zl{observe} operation is only executed based on a dynamic condition on the lengths of two list data structures. Because this condition is not guaranteed to be met, this program cannot execute in bounded memory.

\emph{SLAM}. Every unseparated path has length at most 1. The analysis concludes that the environment map array is not consumed because the model makes random choices that are not guaranteed to cover all the entries of the map. However, manual examination shows that an entry of the map that is never covered by a random choice is 0-consumed by virtue of being never used. Thus, the analysis soundly but imprecisely determines that the $m$-consumed condition fails. 

\subsection{Discussion}

For the \emph{Outlier} and \emph{MTT} benchmarks, even though both fail the $m$-consumed semantic property and therefore are not guaranteed to execute in bounded memory, they will \emph{almost certainly} execute in bounded memory. For example, in \emph{Outlier}, the only way that the memory consumption of the model will increase indefinitely is if a particular random choice always takes one branch, which is a probability-zero event. In general, our semantic properties and analysis implementation reason about the absence of \emph{any} program execution that yields unbounded memory. However, in practice, almost certain bounded-memory execution may also be a useful property of programs.

In general, the analysis can provide a sound guarantee that a program executes with bounded memory. However, as we saw with \emph{SLAM}, it is not always precise enough such that if it rejects a program, then the program must have unbounded memory consumption. For example, it is possible to deliberately construct pathological programs requiring a large number of iterations for the unseparated paths analysis. Remaining limitations on precision include common static analysis challenges such as path sensitivity due to \zl{if} statements and aliasing due to complex data structures. 

When facing conditional branches, the analysis takes a conservative approach that may not utilize all statically available knowledge. Specifically, it cannot determine that certain branches are taken at most once over the entire input stream or that only certain program paths are valid over multiple sequential branches. The analysis also cannot accurately track variables that are stored into complex data structures, meaning it cannot mark them as consumed. We discuss these challenges in greater detail and provide specific examples in \Cref{sec:precision-app}.

\section{Related Work}
\label{sec:related}

\paragraph{Resource Analysis for Probabilistic Programs}

Static resource analysis is capable of automatically determining upper bounds for resources such as time or memory required to execute a probabilistic program. \citet{ngo18} proposed a weakest-precondition approach to determine the expected memory usage of a probabilistic program, which bounds the number of loop iterations executed and number of explicit memory allocation ticks encountered. Our analysis, on the other hand, extends static reasoning to the inherent memory usage of the inference algorithm itself.

\paragraph{Reactive Probabilistic Programming}

\citet{probcc-concur97} first introduced the idea of \emph{reactive} probabilistic programming. They extend a concurrent constraint language with random variables. In contrast, our language is based on a synchronous dataflow model and focus on resource analysis.

\citet{probzelus} developed {\ProbZelus}, a reactive probabilistic programming language which operates over streams of data and supports inference at each stream iteration. It uses an implementation of delayed sampling designed to provide bounded-memory inference for a class of reactive probabilistic programs. However, {\ProbZelus} provides no static guarantee of bounded-memory inference. In this work, we define a language that can be used as a target for the compilation of \ProbZelus and identify the semantic conditions and a static analysis that makes it possible to provide a static guarantee.

\paragraph{Delayed Sampling and Bounded-Memory Inference}

The mechanism of delayed sampling in probabilistic programs was introduced by \citet{murray18delayed_sampling} and implemented in the Anglican and Birch programming languages, neither of which supports inference over streams. Delayed sampling, a form of Sequential Monte Carlo \cite{liu-1998}, can execute in bounded memory because it automates the construction of Rao-Blackwellized particle filters \cite{doucet2000}, a particularly efficient variant of SMC. By comparison, Markov chain Monte Carlo techniques generally cannot execute in bounded memory because they maintain a sample of the full history of program execution, the size of which can grow without bound for a probabilistic stream. Variational inference has extensions that make it amenable to streaming \cite{broderick-2013}, but we are not aware of any probabilistic programming system that makes use of them. 

Other programming languages such as Hakaru \cite{narayanan2016probabilistic} use static program transformations to accomplish the same goal of deferring approximate inference as much as possible. It is unclear if these transformations apply to a streaming context, where dynamic information is necessary to reflect the evolution of the underlying model over many iterations.

\section{Conclusion}
\label{sec:conclusion}

Probabilistic programming has been augmented by constructs that perform inference over unbounded iterations on streams of data. Underlying this programming model is delayed sampling, which combines the benefits of exact inference and the flexibility of sampling. 

In our paper, we introduce the $m$-consumed and unseparated path semantic properties, which show that delayed sampling can execute in bounded memory for reactive probabilistic programs. We present a sound static analysis that verifies these two properties with a type system and an abstract delayed sampling graph. 
To the best of our knowledge, our work is the first to develop a resource analysis for a probabilistic program in relation to its probabilistic programming system's underlying inference algorithm. We hope this work will enable automatic inference mechanisms whose performance is better understood by model developers in probabilistic programming languages.

\begin{acks}
We would like to thank Cambridge Yang, Alex Renda, Jesse Michel, and Ben Sherman, who all provided feedback on drafts of this paper.
This work was supported in part by the MIT-IBM Watson AI Lab and the Office of Naval Research (ONR N00014-17-1-2699).
Any opinions, findings, and
   conclusions or recommendations expressed in this material are those
   of the authors and do not necessarily reflect the views of the
   Office of Naval Research.
\end{acks}

\balance
\bibliography{biblio}

\newpage
\appendix
\section{Ideal Semantics}
\label{sec:ideal-semantics-app}

In this section we present the complete semantics of the deterministic part of \muF in \Cref{fig:muF-sem-full} and the ideal semantics of the probabilistic part in \Cref{fig:muF-psem}.

The probabilistic semantics of \Cref{fig:muF-psem} is a measure-based semantics similar to one presented in~\cite{staton17}.
Given an environment $\gamma$, an expression is interpreted as a measure $\psem{e}_{\gamma}: \Sigma_D \rightarrow [0, 1)$, that is, a function which associates a positive number to each measurable set $U \in \Sigma_D$, where $\Sigma_D$ denotes the $\Sigma$-algebra of the domain of the expression $D$, i.e., the set of measurable sets of possible values.
$\muFsample{v}$ returns the distribution $\sem{v}_{\gamma}$.
$\muFobserve{v_1}{v_2}$ weights execution paths using the likelihood of the observation~$\sem{v_2}_{\gamma}$ w.r.t. the distribution $\sem{v_1}_{\gamma}$ (for a distribution $\mu$ we note $\mu_{\rm{pdf}}$ its probability density function).
Local definitions are interpreted as integration, and we use the Dirac delta measure to interpret deterministic expressions.

\begin{figure}
$$
\begin{array}{@{}l@{\;\;}c@{\;\;}l}
\sem{\muFval{x}{e}}_\gamma &=& \gamma[x \leftarrow \sem{e}_\gamma]
\\[0.5em]

\sem{\muFfun{f}{p}{e}}_\gamma &=& \gamma[f \leftarrow (\fun{v}\sem{e}_{\gamma+[v/p]})]
\\[0.5em]

\sem{d_1\ d_2}_\gamma &=&
  \letin{\gamma_1 = \sem{d_1}_\gamma} \sem{d_2}_{\gamma_1}
\\[0.5em]



\multicolumn{3}{@{}l}{
\sem{\muFval{m}{\muFstream{e_{\textit{init}}}{p_{\mathit{state}}}{p_{\mathit{input}}}{e}}}_\gamma
}
\\&=& \gamma[m \leftarrow \muFstream{e_{\textit{init}}}{p_{\mathit{state}}}{p_{\mathit{input}}}{e}_\gamma]
\\[1em]

\sem{c}_\gamma & = & c
\\[0.5em]

\sem{x}_\gamma & = & \gamma(x)
\\[0.5em]

\sem{\muFpair{v_1}{v_2}}_\gamma & = &
  \muFpair{\sem{v_1}_\gamma}{\sem{v_2}_\gamma}
\\[0.5em]

\sem{\muFop{v}}_\gamma &=& \op(\sem{v}_\gamma)
\\[0.5em]

\sem{\muFapp{f}{v}}_\gamma &=& \gamma(f)(\sem{v}_\gamma)
\\[0.5em]

\sem{\muFletin{p}{e_1}{e_2}}_\gamma &=&
  \letin{v = \sem{e_1}_{\gamma}}\sem{e_2}_{\gamma+[v/p]}
\\[0.5em]

\sem{\muFif{v}{e_1}{e_2}}_\gamma &=&
  \mathit{if} \; \sem{v}_\gamma
  \; \mathit{then} \; \sem{e_1}_\gamma
  \; \mathit{else} \;  \sem{e_2}_\gamma
\\[0.5em]

\sem{\muFinit{m}}_\gamma &=&
  \letin{\muFstream{e_{\mathit{init}}}{p_{\mathit{state}}}{p_{\mathit{input}}}{e}_{\gamma'} = \sem{m}_\gamma} \\&&
  \letin{s_{\mathit{init}} = \sem{e_{\mathit{init}}}_{\gamma'}} \\&&
  (s_{\mathit{init}}, \fun{(s, v)} \sem{e}_{\gamma'+[s/p_{\mathit{state}}, v/p_{\mathit{input}}]})
  \quad \text{if $e$ is deterministic}
  \\[0.5em]

\sem{\muFinit{m}}_\gamma &=&
  \letin{\muFstream{e_{\mathit{init}}}{p_{\mathit{state}}}{p_{\mathit{input}}}{e}_{\gamma'} = \sem{m}_\gamma} \\&&
  \letin{s_{\mathit{init}} = \sem{e_{\mathit{init}}}_{\gamma'}} \\&&
  (s_{\mathit{init}}, \fun{(s, v)} \psem{e}_{\gamma'+[s/p_{\mathit{state}}, v/p_{\mathit{input}}]})
  \quad \text{if $e$ is probabilistic}
  \\[0.5em]


\sem{\muFunfold{x}{v}}_\gamma &=&
  \begin{array}[t]{@{}l}
    \letin{v_{\mathit{state}}, f = \sem{x}_\gamma}\\
    \letin{v_{\mathit{output}}, v_{\mathit{state}}' = f(v_{\mathit{state}}, \sem{v}_\gamma)} \\
    (v_{\mathit{output}}, (v_{\mathit{state}}', f))
  \end{array}
\\[3.em]

\sem{\muFinfer{m}}_\gamma &=&
  \letin{\muFstream{e_{\mathit{init}}}{p_{\mathit{state}}}{p_{\mathit{input}}}{e}_{\gamma'} = \sem{m}_\gamma} \\&&
  \letin{s_{\mathit{init}} = \sem{e_{\mathit{init}}}_{\gamma'}}\,
  (\delta_{s_{\mathit{init}}}, \mathit{infer}(\fun{(s,v)} \psem{e}_{\gamma'+[s/p_{\mathit{state}}, v/p_{\mathit{input}}]})) \\ &&
  \text{where }
  \begin{array}[t]{@{}l}
    \mathit{infer}(f) = \fun{(\sigma, v)}
      \begin{array}[t]{@{}l}
      \letin{\mu = \fun{U} \int_{S} \sigma(ds) f(s, v)(U)}\\
      \letin{\nu = \fun{U} \mu(U) / \mu(\top)}\\
      (\pi_{1*}(\nu), \pi_{2*}(\nu))\\[1em]
    \end{array}
  \end{array}
\\
\end{array}
$$
\caption{Deterministic semantics of \muF. }
\label{fig:muF-sem-full}
\end{figure}

\begin{figure}
$$
\begin{array}{@{}l@{\;\;}c@{\;\;}l}

\psem{v}_\gamma &=&
  \fun{U} \delta_{\sem{v}_\gamma}(U)
\\[0.5em]

\psem{\muFop{v}}_\gamma &=& \fun{U} \delta_{\op(\sem{v}_\gamma)}(U)
\\[0.5em]

\psem{\muFapp{f}{v}}_\gamma &=& \fun{U} \delta_{\gamma(f)(\sem{v}_\gamma)}(U)
\\[0.5em]

\psem{\muFletin{p}{e_1}{e_2}}_\gamma &=&
  \fun{U} \int_T \psem{e_1}_{\gamma}(du) \psem{e_2}_{\gamma+[u/p]}(U)
\\[0.5em]

\psem{\muFif{v}{e_1}{e_2}}_\gamma &=&
  \fun{U}
  \mathit{if} \; \sem{v}_\gamma
  \; \mathit{then} \; \psem{e_1}_\gamma(U)
  \; \mathit{else} \;  \psem{e_2}_\gamma(U)
\\[0.5em]


\psem{\muFunfold{x}{v}}_\gamma &=& \fun{U}
  \begin{array}[t]{@{}l}
    \letin{v_{\mathit{state}}, f = \sem{x}_\gamma}\\
    \letin{\mu = f(v_{\mathit{state}}, \sem{v}_\gamma)} \\
    \int \mu(dv_{\mathit{output}}, dv_{\mathit{state}}') \delta_{(v_{\mathit{output}},(v_{\mathit{state}}', f))}(U)
  \end{array}
\\[3.em]

\psem{\muFsample{v}}_\gamma &=& \fun{U} \sem{v}_\gamma(U)
\\[0.5em]

\psem{\muFobserve{v_1}{v_2}}_\gamma &=&
  \fun{U}
  \letin{\mu = \sem{v_1}_\gamma} \mu_{\textrm{pdf}}(\sem{v_2}_\gamma) * \delta_{\mathtt{()}}(U)
\\[0.5em]

\end{array}
$$
\caption{Probabilistic semantics of \muF. }
\label{fig:muF-psem}
\end{figure}


\newcommand{\mufty}{T}
\newcommand{\muftydet}{T_D}
\newcommand{\muftyprob}{T_P}
\newcommand{\muftydistr}{\texttt{distr}}
\newcommand{\muftystreamdet}{\texttt{dstream}}
\newcommand{\muftystreamprob}{\texttt{pstream}}
\newcommand{\muftyjudgdet}[3]{{{#1} \vdash_\mathit{det} {#2} : {#3}}}
\newcommand{\muftyjudgprog}[3]{{{#1} \vdash_\mathit{decl} {#2} : {#3}}}
\newcommand{\muftyjudgprob}[3]{{{#1} \vdash_\mathit{prob} {#2} : {#3}}}
\newcommand{\muftyjudgparam}[3]{{{#1} \vdash_\mathit{k} {#2} : {#3}}}
\newcommand{\muftystreamfndet}{\texttt{dstreamfn}}
\newcommand{\muftystreamfnprob}{\texttt{pstreamfn}}
\newcommand{\muftyinferstate}{\texttt{istate}}
\newcommand{\muftyinferout}{\texttt{result}}
\newcommand{\muftymeasurable}[1]{\textsf{measurable}({#1})}

\section{Core types in $\muF{}$}
\label{sec:muF-types}

This section describes a type system for $\muF{}$ programs. 
All programs we consider in this work must type check according to this system.
The type system ensures that if an expression $e$ is given a probabilistic typing judgment $\muftyjudgprob{\Gamma}{e}{T}$ (which means that $e$ will be evaluated using its probabilistic semantics $\psem{e}$ rather than its deterministic semantics $\sem{e}$), then its type $T$ is a measurable space that does not include nonmeasurable objects such as functions.
The type system also prohibits nested inference.

The types of $\muF{}$ are unit, Booleans, reals, functions, and pairs, as well as probability distributions, and deterministic and probabilistic stream functions and stream instances.
\begin{align*}
\mufty & ::= \texttt{unit} \; \mid \; \texttt{bool} \; \mid \; \texttt{real} \; \mid \; \mufty \rightarrow \mufty \; \mid \; \mufty \times \mufty \; \mid \; \muftydistr \; \mufty \\
    & \; \mid \; \muftystreamfndet (\mufty, \mufty) \; \mid \; \muftystreamdet (\mufty, \mufty) \; \mid \; \muftystreamfnprob (\mufty, \mufty) \; \mid \; \muftystreamprob (\mufty, \mufty)
\end{align*}
Only a subset of these types may act as the support of probability distributions, denoted by the judgment $\muftymeasurable{T}$. These exclude function and stream types:
\begin{mathpar}
\inferrule{\vphantom{\texttt{unit}}}{\muftymeasurable{\texttt{unit}}}

\inferrule{\vphantom{\texttt{bool}}}{\muftymeasurable{\texttt{bool}}}

\inferrule{\vphantom{\texttt{real}}}{\muftymeasurable{\texttt{real}}}

\inferrule{\muftymeasurable{\mufty_1} \\ \muftymeasurable{\mufty_2}}{\muftymeasurable{\mufty_1 \times \mufty_2}}

\inferrule{\muftymeasurable{\mufty}}{\muftymeasurable{\muftydistr \; \mufty}}
\end{mathpar}

We present the full type system of $\muF{}$ in~\Cref{fig:muF-types,fig:muF-types-program}.

\begin{figure}
\begin{mathpar}
\inferrule{
    \muftyjudgdet{\Gamma}{e}{T}
}{
    \muftyjudgprog{\Gamma}{\muFval{p}{e}}{\Gamma, p : T}
}

\inferrule{
    \muftyjudgdet{\Gamma, p : T}{e}{T'} 
} {
    \muftyjudgprog{\Gamma}{\muFfun{f}{p}{e}}{\Gamma, p : T \rightarrow T'}
}

\inferrule{
    \muftyjudgprog{\Gamma}{d_1}{\Gamma'}\\
    \muftyjudgprog{\Gamma'}{d_2}{\Gamma''}
} {
    \muftyjudgprog{\Gamma}{d_1 \; d_2}{\Gamma''}
}

\inferrule{
    \muftyjudgdet{\Gamma}{e_\textit{init}}{T_\textit{state}}\\
    \muftyjudgdet{\Gamma, (p_\textit{state}, p_\textit{input}) : T_\textit{state} \times T_\textit{input}}{e_\textit{step}}{T_\textit{state} \times T_\textit{out}} \\
} {
    \muftyjudgprog{\Gamma}{\muFval{m}{\muFstream{e_\textit{init}}{p_\textit{state}}{p_\textit{input}}{e_\textit{step}}}}{\Gamma, m : \muftystreamfndet (T_\textit{input}, T_\textit{out})}
}

\inferrule{
    \muftyjudgprob{\Gamma}{e_\textit{init}}{T_\textit{state}}\\
    \muftyjudgprob{\Gamma, (p_\textit{state}, p_\textit{input}) : T_\textit{state} \times T_\textit{input}}{e_\textit{step}}{T_\textit{state} \times T_\textit{out}} \\
} {
    \muftyjudgprog{\Gamma}{\muFval{m}{\muFstream{e_\textit{init}}{p_\textit{state}}{p_\textit{input}}{e_\textit{step}}}}{\Gamma, m : \muftystreamfnprob (T_\textit{input}, T_\textit{out})}
}
\end{mathpar}
\caption{Typing rules for programs in $\muF$. The judgment $\muftyjudgprog{\Gamma}{d}{\Gamma'}$ means that the $\muF$ declaration $d$, when typed under the typing context $\Gamma$, produces the typing context $\Gamma'$.}
\label{fig:muF-types-program}
\end{figure}

\begin{figure}
\begin{mathpar}
\inferrule{ } {
    \muftyjudgdet{\Gamma}{\texttt{()}}{\texttt{unit}}
}

\inferrule {
    c \in \{\texttt{true}, \texttt{false}\}
} {
    \muftyjudgdet{\Gamma}{c}{\texttt{bool}}
}

\inferrule {
    c \in \mathbb{R}
} {
    \muftyjudgdet{\Gamma}{c}{\texttt{real}}
}

\inferrule {
    \muftyjudgdet{\Gamma}{e_1}{T_1}\\
    \muftyjudgdet{\Gamma}{e_2}{T_2}\\
} {
    \muftyjudgdet{\Gamma}{\texttt{(}e_1 \texttt{,} e_2 \texttt{)}}{T_1 \times T_2}
}

\inferrule{
    \muftyjudgparam{\Gamma, p_1 : T_1, p_2 : T_2}{e}{T'}
} {
    \muftyjudgparam{\Gamma, (p_1,p_2) : T_1 \times T_2}{e}{T'}
}

\inferrule{ } {
    \muftyjudgparam{\Gamma, x : T}{x}{T}
}

\inferrule{
    \muftyjudgdet{\Gamma}{e}{T}\\
    \muftyjudgparam{\Gamma}{f}{T \rightarrow T'}
} {
    \muftyjudgparam{\Gamma}{\muFapp{f}{e}}{T'}
}

\inferrule{
    \muftyjudgdet{\Gamma}{e_1}{\texttt{bool}}\\
    \muftyjudgparam{\Gamma}{e_2}{T}\\
    \muftyjudgparam{\Gamma}{e_3}{T}
} {
    \muftyjudgparam{\Gamma}{\muFif{e_1}{e_2}{e_3}}{T}
}

\inferrule{
    \muftyjudgparam{\Gamma}{e_1}{T_1}\\
    \muftyjudgparam{\Gamma, p : T_1}{e_2}{T_2}
} {
    \muftyjudgparam{\Gamma}{\muFletin{p}{e_1}{e_2}}{T_2}
}

\inferrule{
    \muftyjudgdet{\Gamma}{m}{\muftystreamfndet(T_\textit{input}, T_\textit{out})}
}{
    \muftyjudgdet{\Gamma}{\muFinit{m}}{\muftystreamdet(T_\textit{input}, T_\textit{out})}
}

\inferrule{
    \muftyjudgdet{\Gamma}{m}{\muftystreamfnprob(T_\textit{input}, T_\textit{out})}
}{
    \muftyjudgdet{\Gamma}{\muFinfer{m}}{\muftystreamprob(T_\textit{input}, T_\textit{out})}
}

\inferrule{
    \muftyjudgdet{\Gamma}{e_1}{\muftystreamdet(T_\textit{input}, T_\textit{out})} \\
    \muftyjudgdet{\Gamma}{e_2}{T_\textit{input}}
}{
    \muftyjudgparam{\Gamma}{\muFunfold{e_1}{e_2}}{T_\textit{out} \times \muftystreamdet(T_\textit{input}, T_\textit{out})}
}

\inferrule{
    \muftyjudgdet{\Gamma}{e_1}{\muftystreamprob(T_\textit{input}, T_\textit{out})} \\
    \muftyjudgdet{\Gamma}{e_2}{T_\textit{input}}
}{
    \muftyjudgparam{\Gamma}{\muFunfold{e_1}{e_2}}{\muftydistr \; T_\textit{out} \times \muftystreamprob(T_\textit{input}, T_\textit{out})}
}

\inferrule{ 
    \muftyjudgdet{\Gamma}{e}{T} \\
    \muftymeasurable{T}
} {
    \muftyjudgprob{\Gamma}{e}{T}
}

\inferrule{
    \muftyjudgdet{\Gamma}{e}{\muftydistr \; \mufty}
}{
    \muftyjudgprob{\Gamma}{\muFsample{e}}{\mufty}
}

\inferrule{
    \muftyjudgdet{\Gamma}{e_1}{\muftydistr \; \mufty} \\
    \muftyjudgdet{\Gamma}{e_2}{\mufty}
}{
    \muftyjudgprob{\Gamma}{\muFobserve{e_1}{e_2}}{\texttt{unit}}
}
\end{mathpar}
\caption{Deterministic and probabilistic type systems for \muF{}. The typing judgment $\muftyjudgdet{\Gamma}{e}{T}$ means that the $\muF$ expression $e$ under the context $\Gamma$ has the deterministic type $T$. The judgment $\muftyjudgprob{\Gamma}{e}{T}$ means that the $\muF$ expression $e$ under context $\Gamma$ has the probabilistic type $T$. The judgment $\muftyjudgparam{\Gamma}{e}{T}$ stands for either the deterministic or the probabilistic judgment, where $k$ is instantiated to be $det$ or $prob$. These rules state that $\zl{sample}$ and $\zl{observe}$ can only be used inside the body of a probabilistic stream.}
\label{fig:muF-types}
\end{figure}


\section{Definition of $\mathit{graft}$}
\label{sec:graft-app}

\newcommand{\dsgraft}{\ensuremath{\mathit{graft}}}
\newcommand{\dsprune}{\ensuremath{\mathit{prune}}}

In this section, we review the definition of $\dsgraft$ from \citet{murray18delayed_sampling}.
\subsection{Preliminaries}
This definition makes use of an alternative type of marginalized node that maintains its own marginal distribution as well as a conditional distribution that relates the marginalized node to its unique marginalized child.
We use the notation $\stmarginalized(\mu_\mathit{marg}, \mu_\mathit{cond})$ to refer to such a marginalized node with marginal distribution $\mu_\mathit{marg}$ and conditional distribution $\mu_\mathit{cond}$.
We use the notation $s = \stmarginalized(\_)$ to mean that the node state $s$ is a marginalized node of any type.
The two types of marginalized nodes only differ in the distributions they store, and have the same reachability and memory consumption properties.

\citet{murray18delayed_sampling} defines invariants of delayed sampling runtimes. 
Namely, it specifies that delayed sampling maintains that (1) all nodes in the delayed sampling graph have at most one parent, and (2) all marginalized nodes in the graph have at most one marginalized or realized child.
In the following definitions, we use the notation $\mathit{parent}(X, E)$ to mean a function that returns the unique parent of $X$ in the edge set $E$.
We also use the notation $\mathit{child}(X, E)$ to mean a function that returns the unique realized or marginalized child of the marginalized node $X$ in the edge set $E$.

\subsection{Definitions}

We define the $\mathit{graft}$ function as follows. When called on an initialized node, $\mathit{graft}$ recursively marginalizes every initialized ancestor of the given node.
This means that it performs integration to incorporate parent information into the distributions of each node in the initialized chain. When called on a marginalized node, $\mathit{graft}$ calls the $\mathit{prune}$ function.

$$
\begin{array}{lcl}
  \dsgraft(X, g) & = & \letin{(V, E, q) = g}\\&&
    \mathit{if} \; q(X) = \stinitialized(\mu) \; \mathit{then}
    \\&&
       \quad\begin{array}{@{}l}
       \letin{X_\mathit{par} = \mathit{parent}(X, E)}\\
       \mathit{let}\ \mu_\mathit{prior}, g' = \\
         \quad\begin{array}{@{}l}
         \mathit{if} \;
           \begin{array}[t]{@{}l@{}}
            q(X_\mathit{par}) = \stmarginalized(\mu_\mathit{par}) \; \mathit{or}\\
            q(X_\mathit{par}) = \stinitialized(\mu_\mathit{par})
           \end{array}\\
         \mathit{then} \;
           \begin{array}[t]{@{}l}
             \letin{(V'', E'', q'') = \dsgraft(X_\mathit{par}, g)} \\
             \letin{\stmarginalized(\mu_\mathit{par}) = q''(X_\mathit{par})}\\
             \mu_\mathit{par}, (V'', E'', q''[X_\mathit{par} \leftarrow \stmarginalized(\mu_\mathit{par}, \mu)])
            \end{array}
         \\
         \mathit{else} \; \mathit{if} \; q(X_\mathit{par}) = \strealized(v) \; \mathit{then} \; \delta(v), (V, E - (X, X_\mathit{par}), q)
         \end{array}\\
       \mathit{in} \\
       \letin{\mu', (V'', E'', q'') = \int \mu \; \mathrm{d}\mu_\mathit{prior}, g'}\\
       (V'', E'', q''[X \leftarrow \stmarginalized(\mu')])
       \end{array}
    \\&&
    \mathit{else} \; \mathit{if} q(X) = \stmarginalized(\mu, \mu_\mathit{child}) \; \mathit{then} \; \dsprune(X, g)
    \\&&
    \mathit{else} \; g
\end{array}
$$

We define the $\mathit{prune}$ function as follows. 
When called on a marginalized node with a marginalized or realized child, the function first recursively prunes that child if the child itself is marginalized. If the node is marginalized, it samples a value for that node and then conditions the current node on the child taking on that value. If the child node is realized, the function proceeds to immediately condition the current node on the child node's value.

In either case, the conditioning proceeds as follows.
The $\mathit{prune}$ function first extracts probability density functions from the relevant measures using the $\mathit{pdf}$ function.
It then follows Bayes' rule, multiplying the prior and conditional density functions and normalizing the result with the $\mathit{normalize}$ function.
It finally updates the marginal distribution of the given node and removes the edge connecting the node to its child.

$$
\begin{array}{lcl}
    \dsprune(X, g) & = & \letin{(V,E,q) = g} \\
    &&
     \mathit{if} \; q(X) = \stmarginalized(\mu_X, \mu) \; \mathit{then} \\
    && \quad\begin{array}{l@{}}
         \letin{X_\mathit{child} = \mathit{child}(X, E)}\\
         \letin{g' = \dsprune(X_\mathit{child}, g)}\\
         \mathit{if} \; q(X_\mathit{child}) = \stmarginalized(\mu_\mathit{child}) \; \mathit{then}\\
         \quad\begin{array}{l}
           \letin{v, (V'', E'', q'') = \dsvalue(X_\mathit{child}, g')} \\
           \letin{p_X, p_{\mathit{child} | X} = \mathit{pdf}(\mu_X), \mathit{pdf}(\mu)}\\
           \letin{\mu_X' = \mathit{normalize}(\lambda x. p_X(x) * p_{\mathit{child} | X}(v | x))}\\
           \letin{q''' = q''[X_\mathit{child} \leftarrow \strealized(v), X \leftarrow \stmarginalized(\mu_X')]}\\
           (V'', E'' - (X, X_\mathit{child}), q''')
         \end{array}\\
         \mathit{else} \; \mathit{if} \; q(X_\mathit{child}) = \strealized(v) \; \mathit{then} \\
         \quad\begin{array}{l}
           \letin{v, (V'', E'', q'') = g'} \\
           \letin{p_X, p_{\mathit{child} | X} = \mathit{pdf}(\mu_X), \mathit{pdf}(\mu)}\\
           \letin{\mu_X' = \mathit{normalize}(\lambda x. p_X(x) * p_{\mathit{child} | X}(v | x))}\\
           \letin{q''' = q''[X \leftarrow \stmarginalized(\mu_X')]}\\
           (V'', E'' - (X_\mathit{child}, X), q''')\\
         \end{array}\\
         \mathit{else} \; g
     \end{array}\\
    &&
     \mathit{else} \; g
\end{array}
$$


\section{Complete Analysis Type System}
\label{sec:types-app}

The following is the complete definition of the typing judgment $\hastypealpha{\ctx}{\graph}{e}{\type}{\graph'}$ describing the types and abstract graph transitions of expressions.
\begin{mathpar}
    \inferrule{ }{\vhastypealpha{\ctx}{c}{(\emptyset,\emptyset)}}

    \inferrule{ }{\vhastypealpha{\ctx, x : \type}{x}{\type}}

    \inferrule{ }{\vhastypealpha{\ctx, m : (\type_{\mathit{init}}, \stepfn{p_{\mathit{state}}}{p_{\mathit{in}}}{\ctx_e}{e})}{m}{(\type_{\mathit{init}}, \stepfn{p_{\mathit{state}}}{p_{\mathit{in}}}{\ctx_e}{e})}}

    \inferrule{\vhastypealpha{\ctx}{v}{r} \and \freshvar{X}{\graph}}{\hastypealpha{\ctx}{\graph}{\muFsample{v}}{(\{X\},\{X\})}{\assumej{\alpha}{X}{r}{\graph}}}

    \inferrule{\hastypealpha{\ctx}{\graph}{\muFsample{v_1}}{(\{X\},\{X\})}{\graph'} \and \vhastypealpha{\ctx}{v_2}{r_2}}{\hastypealpha{\ctx}{\graph}{\muFobserve{v_1}{v_2}}{()}{\observej{\alpha}{X}{r_2}{\valuej{\alpha}{r_2}{\graph'}}}}

    \inferrule{\vhastypealpha{\ctx}{v}{\type} \and \type \rfold r}{\vhastypealpha{\ctx}{\muFop{v}}{r}}

    \inferrule{\hastypealpha{\ctx}{\graph}{e}{\type}{\graph'} \and \hastypealpha{\ctx, p : \type}{\graph'}{e'}{\type'}{\graph''}}{\hastypealpha{\ctx}{\graph}{\muFletin{p}{e}{e'}}{\type'}{\graph''}}

    \inferrule{\vhastypealpha{\ctx}{v_1}{\type_1} \and \vhastypealpha{\ctx}{v_2}{\type_2}}{\vhastypealpha{\ctx}{\muFpair{v_1}{v_2}}{\type_1 \times \type_2}}

    \inferrule{\vhastypealpha{\ctx}{v}{r} \and \graph' = \valuej{\alpha}{r}{\graph} \and \hastypealpha{\ctx}{\graph'}{e_1}{\type_1}{\graph_1} \and \hastypealpha{\ctx}{\graph'}{e_2}{\type_2}{\graph_2}}{\hastypealpha{\ctx}{\graph}{\muFif{v}{e_1}{e_2}}{\type_1 \sqcup \type_2}{\graph_1 \sqcup_\alpha \graph_2}}

    \inferrule{\vhastypedet{p}{\type} \\ \vhastypealpha{\ctx, p : \type}{e}{\type'}}{\vhastypealpha{\ctx}{\texttt{fun } p \texttt{ -> } e}{\type \to \type'}}

    \inferrule{\vhastypealpha{\ctx}{f}{\type \to \type'} \and \vhastypealpha{\ctx}{v}{\type}}{\vhastypealpha{\ctx}{f(v)}{\type'}}

    \inferrule{\vhastypealpha{\ctx}{m}{(\type, s)}}{\vhastypealpha{\ctx}{\muFinit{m}}{\pstream{\type}{s}}}

    \inferrule{\ctx \vdash_{mc} m\ \bounded \\ \ctx \vdash_{up} m\ \bounded}{\vhastypealpha{\ctx}{\muFinfer{m}}{\bounded}}

    \inferrule{\vhastypealpha{\ctx}{x}{\pstream{\type}{\stepfn{p_{\mathit{state}}}{p_{\mathit{in}}}{\ctx_e}{e}}} \\ \vhastypealpha{\ctx}{v}{\type_{\mathit{in}}} \\ \hastypealpha{\ctx_e, p_{\mathit{state}} : \type, p_{\mathit{in}} : \type_{\mathit{in}}}{\graph}{e}{\type' \times \type_{\mathit{out}}}{\graph'}}{\hastypealpha{\ctx}{\graph}{\muFunfold{x}{v}}{\type_{\mathit{out}} \times \pstream{\type'}{\stepfn{p_{\mathit{state}}}{p_{\mathit{in}}}{\ctx_e}{e}}}{\graph'}}

    \inferrule{\vhastypealpha{\ctx}{x}{\bounded} \and \vhastypealpha{\ctx}{v}{\type} \and \type \rfold (\emptyset, \emptyset)}{\vhastypealpha{\ctx}{\muFunfold{x}{v}}{(\emptyset,\emptyset) \times \bounded}}
\end{mathpar}
\newcommand{\wellformed}[2]{{#1} \vdash_\alpha {#2} :: \mathit{program}}

\muF{} programs consist of a series of value, function, and stream function declarations. Thus, we also define a top-level judgment $\wellformed{\ctx}{d}$ that states that a \muF{} program $d$ contains declarations that are all well-formed. This judgment is defined as follows:
\begin{mathpar}
    \inferrule{ }{\wellformed{\ctx}{\epsilon}}

    \inferrule{
        \hastypealpha{\ctx}{\graph}{e}{\type}{\graph'} \\
        \wellformed{\ctx, p : t}{d}
    }{\wellformed{\ctx}{\muFval{p}{e}; d}}

    \inferrule{
        \hastypealpha{\ctx}{\graph}{\texttt{fun } p \texttt{ -> } e}{\type}{\graph'} \\
        \wellformed{\ctx, f : t}{d}
    }{\wellformed{\ctx}{\muFfun{f}{p}{e}; d}}

    \inferrule{
        \vhastypealpha{\ctx}{\muFstream{e'}{p_{\mathit{state}}}{p_{\mathit{in}}}{e}}{(\type_{\mathit{init}}, \stepfn{p_{\mathit{state}}}{p_{\mathit{in}}}{\ctx}{e})} \\
        \wellformed{\ctx, m : (\type_{\mathit{init}}, \stepfn{p_{\mathit{state}}}{p_{\mathit{in}}}{\ctx}{e})}{d}
    }{\wellformed{\ctx}{\muFval{m}{\muFstream{e'}{p_{\mathit{state}}}{p_{\mathit{in}}}{e}}; d}}
\end{mathpar}
where if $d$ is empty (i.e. all streams including \texttt{main} are valid) the judgment holds trivially.


\section{Soundness}
\label{app:soundness}
\subsection{Executions}
\label{subsec:execution}

During the execution of a program, the only constructs that can dynamically allocate memory are \zl{sample} and \zl{observe} which add a new node to the delayed sampling graph using the $\dsassume$ operation.
These two probabilistic constructs can only be used in a model, i.e., the argument of the \zl{infer} operator.
We thus focus on the memory footprint of \zl{infer}'s transition function.

The execution of the transition function $\dsinfer$ of \zl{infer} comprises three steps~(see \Cref{sec:ds_op_sem}): (1)~draw a set of particles, i.e.,  pairs (state, graph), (2)~execute the model for each particle, (3) extract the distributions of state and outputs.
The only operation that can dynamically allocate memory is the second one, where the delayed sampling graph can be altered.

At iteration $n$, for each particle, the current pair (state, graph) is obtained from a succession of application of the model transition function from the initial state~$s_0$ and an empty graph~$g_0 =\emptyset$ (step (1) in the definition of $\dsinfer$ can only drop some execution paths).
We call this sequence $(s_0, g_0), (s_1, g_1), \dots $ an \emph{execution} of the model.
The following properties states that if only bounded-memory execution are possible, then the $\dsinfer$ function executes in bounded-memory.

\begin{lemma}[Execution Sufficiency]
    \label{lem:executions}
For all stream functions $m$ and environments $\gamma$,
let \newline $\muFstream{e_{\mathit{init}}}{p_{\mathit{state}}}{p_{\mathit{input}}}{e}_{\gamma'} = \sem{m}_\gamma$ and let $s_m = \sem{e_{\mathit{init}}}_{\gamma'}$ and let $f_m = \lambda (s, v). \; \psem{e}_{\gamma+[s/p_{\mathit{state}},v/p_{\mathit{input}}]}$ and let $s_i, f_i = \sem{\zl{infer(}m\zl{)}}_{\gamma}$.
We say that $f_i$ produces a sequence of distributions $(\mu_n)_{n \in \mathbb{N}}$ given an input sequence $(i_n)_{n \in \mathbb{N}}$ if $f_i(\mu_n, i_n) = (\omega_n, \mu_{n+1})$ for some sequence of output distributions $(\omega_n)_{n \in \mathbb{N}}$.
Similarly, we say $f_m$ produces the execution $(g_n, s_n)_{n \in \mathbb{N}}$ if 
$f_m(s_n, i_n)(g_n, 1) = ((o_n, s_{n+1}), w_n, g_{n+1})$ for some sequences of outputs $(o_n)_{n \in \mathbb{N}}$ and weights $(w_n)_{n \in \mathbb{N}}$.

The lemma states that for all input sequences $(i_n)_{n \in \mathbb{N}}$, if $f_i$ produces the sequence $(\mu_n)_{n \in \mathbb{N}}$, then for all $n$ and $(g_n, s_n) \in \text{support}(\mu_n)$, there exists an execution $(g'_{n}, s'_{n})_{n \in \mathbb{N}}$ such that $g'_n = g_n$, $s'_n = s_n$, and $f_m$ produces $(g'_{n}, s'_{n})_{n \in \mathbb{N}}$.
\label{prop:executions}

\end{lemma}
\begin{proof}
Proceed by induction on $n$.
If $n = 0$, the distribution $\mu_0$ is obtained by the execution of~$f_m$ by each particle on the initial state and the empty graph. So the support of $\mu_0$ is obtained by the execution of~$f_m$.
If $n > 0$, by definition of $\mathit{infer}$, each pair $(g_n,s_n)$ from the support of the distribution $\mu_n$ is obtained by the application of $f_m$ on $(g_{n-1},s_{n-1})$ drawn from the distribution $\mu_{n-1}$.
By application of the induction hypothesis, $(g_i,s_i)_{0 \le i < n}$ is an execution produced by $f_m$, and therefore $(g_i,s_i)_{0 \le i \le n}$ is also produced by $f_m$.
\end{proof}

\begin{corollary}
If all executions of $f_m$ are bounded-memory, then for any input sequence $(i_n)_{n \in \mathbb{N}}$ and any execution $(g_n, s_n)_{n \in \mathbb{N}}$ such that for all $n$, every  $(g_n, s_n) \in \text{support}(\mu_n)$, $(g_n, s_n)_{n \in \mathbb{N}}$ has bounded memory when $\mu_n$ is produced by $f_i$.
\label{cor:executions}
\end{corollary}

\newcommand{\vmap}{\ell}
\subsection{Type Soundness}
\label{subsec:typesapp}

In this section, we show that the type system is sound.
We first define the $\vDash$ relations referenced in \Cref{subsec:soundness}.
We then prove the soundness theorems stated in \Cref{subsec:soundness}.

\paragraph{Variable Mappings} Both delayed sampling and the type system use a set of fresh variable names to label random variables. 
Because the type system and the delayed sampling execution may each use a different name for conceptually the same random variable, we define an association that maps between these namespaces. 
We use the notation $\vmap$ to refer to a function that maps a delayed sampling variable to a type system variable, and $\vmap_*$ to the extension of $\vmap$ to sets: $\vmap_*(\hat{X}) = \{ \ell(X) \mid X \in \hat{X} \}$.

\subsubsection{Entailment}

Here we establish what it means for a value to entail a type.
A value entails a type if the type accurately captures the random variables the value could refer to, as well as the shape of the value (i.e.\ whether the value is a scalar, a pair, or a stream function).
Because step function types include type contexts, we also establish what it means for an environment to entail a type context.

A stream value entails $\bounded$ if it produces a sequence of states in which every delayed sampling graph is bounded. We formalize this as follows.
Given a sequence of inputs $(i_n)_{n \in \mathbb{N}}$ and an initial state $s_0$, we say a stream function $f$ produces the sequence of state $(s_n)_{n \in \mathbb{N}}$ on $(s_0, i)$, if $f(i_n, s_n) = (o_n, s_{n+1})$ for some output sequence $(o_n)_{n \in \mathbb{N}}$.
We say $s$, is bounded if every sequence delayed sampling graphs contained in $s$ is low-level bounded-memory.

\begin{definition}[Type Entailment]
    \label{def:type-entail}
    A value $\progvar$ entails a type $\type$, written $\progvar \vDash^\vmap \type$, under the following circumstances:
\begin{align*}
    c \vDash^\vmap (\emptyset, \mathit{ub}) \\
    \dsnode \vDash^\vmap (\mathit{lb}, \mathit{ub}) \iff &  \; \mathit{lb} \subseteq \{\vmap(\dsnode)\} \subseteq \mathit{ub}\\
    \dsapp{\op}{\progvar} \vDash^\vmap (\mathit{lb}, \mathit{ub}) \iff &  \; \mathit{lb} \subseteq \vmap_*(\freerandvars(\progvar)) \subseteq \mathit{ub}\\
    \muFpair{\progvar_1}{\progvar_2} \vDash^\vmap \type_1 \times \type_2 \iff &  \; \progvar_1 \vDash^\vmap \type_1 \; \mathrm{and} \; \progvar_2 \vDash^\vmap \type_2\\
    \muFstream{e_{\mathit{init}}}{p_{\mathit{in}}}{p_{\mathit{state}}}{e_{\mathit{state}}}_{\gamma_e} \\
    \vDash^\vmap (t_{\mathit{init}}, \stepfn{p_{\mathit{in}}}{p_{\mathit{state}}}{\ctx_e}{e_{\mathit{state}}}) \iff &  e_{\mathit{init}} \vDash^\vmap t_{\mathit{init}} \wedge \gamma_e \vDash^\vmap \ctx_e  \\
    (s_0, f) \vDash^\vmap  \pstream{\type_{\mathit{init}}}{S} \iff &  \;  s_0 \vDash^\vmap \type_{\mathit{init}} \; \text{and} \; f \vDash^\vmap S \\
    f \vDash^\vmap \stepfn{p_{\mathit{in}}}{p_{\mathit{state}}}{\ctx_e}{e} \iff &  \; \exists \gamma. \; \gamma \vDash^\vmap \ctx_e \; \text{and} \; f = \lambda (s, v). \; \psem{e}_{\gamma+[s/p_{\mathit{state}},v/p_{\mathit{in}}]} \\
    \gamma \vDash^\vmap \ctx \iff &  \; \forall x : \type \in \ctx. \; \gamma(x) = \progvar \; \text{s.t.} \; \progvar \vDash^\vmap \type\\
    (s_0, f) \vDash \bounded \iff &  \forall \mathit{i}. \; f \; \text{produces} \; s \; \text{on} \; (s_0, i) \\
    & \Rightarrow  s \; \text{is bounded}
\end{align*}
\end{definition}

We further define a version of type entailment that only applies to a restricted set of variables.

\begin{definition}[Restricted Type Entailment]
    A value $\progvar$ entails a type $\type$ -- restricted to the variable set $\hat{X}$, written $\progvar \vDash^\vmap_{\hat{X}} \type$, under the following circumstances:
    \begin{align*}
    c \vDash^\vmap_{\hat{X}} (\emptyset, \mathit{ub}) \\
    \dsnode \vDash^\vmap_{\hat{X}} (\mathit{lb}, \mathit{ub}) \iff & \; \dsnode \in \hat{X} \Rightarrow \mathit{lb} \subseteq \{\vmap(\dsnode)\} \subseteq \mathit{ub}\\
    \dsapp{\op}{\progvar} \vDash^\vmap_{\hat{X}} (\mathit{lb}, \mathit{ub}) \iff &  \; \dsnode \in \hat{X} \Rightarrow \mathit{lb} \subseteq \vmap_*(\freerandvars(\progvar)) \subseteq \mathit{ub}\\
    \end{align*}
    The rules for any other values are similar to those in \Cref{def:type-entail}, but pass the set $\hat{X}$ through unchanged for recursive definitions.
\end{definition}

The fold operation $\rfold$ is designed to generate a scalar type that encapsulates the free variables of a value while disregarding its shape.

\begin{lemma}[Fold Entailment]
    If $\progvar \vDash^\vmap \type$ and $\type \rfold (\mathit{lb}, \mathit{ub})$, then $\mathit{lb} \subseteq \vmap_*(\freerandvars(\progvar)) \subseteq \mathit{ub}$.
\end{lemma}

A traced graph entails an $m$-consumed abstract graph if the abstract graph soundly approximates the variables that are not used.

\begin{definition}[$m$-consumed Graph Entailment]
    A traced graph  $(g, \tau)$ {\em entails} an $m$-consumed abstract graph $\graph$, written $(g, \tau) \vDash_{\mathit{mc}}^\vmap \graph$, if for every variable $X$ not in $\graph.\mathit{in} \setminus \graph.\mathit{con}$ and for every $X_0$ such that $\vmap(X_0) = X$, $X_0$ is used in $\tau$.
\end{definition}

A traced graph entails an unseparated-path abstract graph if the path function soundly approximates the unseparated paths in the traced graph and the separator set soundly approximates the set of variables that are $\dsobserve$d or $\dsvalue$d.

\begin{definition}[Unseparated Path Graph Entailment]
    A graph $(g, \trace)$ {\em entails} an unseparated-path abstract graph $\graph$, written $(g, \trace) \vDash_{\mathit{up}}^\vmap \graph$ if for every $\dsnode_1$, $\dsnode_2$ that are referenced in $\trace$, $\graph.p(\vmap(\dsnode_1),\vmap(\dsnode_2))$ is at least the length of the unseparated path between $\dsnode_1$ and $\dsnode_2$ in $\trace$, and, for all $\dsnode$ referenced in $\trace$, $\graph.sep(\vmap(\dsnode))$ is only true if $\dsnode$ is a separator in $\trace$.
\end{definition}

\paragraph{Entailment from \Cref{subsec:soundness}} Here, we define the entailment relations that are referenced in \Cref{subsec:soundness}. These definitions are defined in terms of the relevant definitions in this section with the variable map $\vmap$ existentially quantified:
\begin{align*}
v \vDash_\alpha \type & \iff \exists \vmap. \; v \vDash^\vmap \type\\
\gamma \vDash_\alpha \Gamma & \iff \exists \vmap. \; \gamma \vDash^\vmap \Gamma\\
v, (g, \trace) \vDash_\alpha \type, \graph & \iff \exists \vmap. \; %
    v \vDash^\vmap t \wedge (g, \trace) \vDash_\alpha^\vmap \graph\\
    \gamma, (g, \trace) \vDash_\alpha \Gamma, \graph & \iff \exists \vmap. \; \gamma \vDash^\vmap \Gamma \wedge (g, \trace) \vDash_\alpha^\vmap \graph
\end{align*}
We further extend these definitions to incorporate a restricted variable set $\hat{X}$.
\begin{align*}
v \vDash_{\alpha, \hat{X}} \type & \iff \exists \vmap. \; v \vDash^\vmap_{\hat{X}} \type\\
\gamma \vDash_{\alpha, \hat{X}} \Gamma & \iff \exists \vmap. \; \gamma \vDash^\vmap_{\hat{X}} \Gamma\\
v, (g, \trace) \vDash_{\alpha, \hat{X}} \type, \graph & \iff \exists \vmap. \; %
    v \vDash^\vmap_{\hat{X}} t \wedge (g, \trace) \vDash_\alpha^\vmap \graph\\
    \gamma, (g, \trace) \vDash_{\alpha, \hat{X}} \Gamma, \graph & \iff \exists \vmap. \; \gamma \vDash^\vmap_{\hat{X}} \Gamma \wedge (g, \trace) \vDash_\alpha^\vmap \graph
\end{align*}

\subsubsection{Soundness Theorems}

\newtheorem*{specialsound}{Special Case of \Cref{thm:mc-up-sound}}
\newtheorem*{restatesound}{Restatement of \Cref{thm:ana-sound}}

An $m$-consumed type judgment is sound if it abstracts the $m$-consumed property of the semantics according to the entailment relations.

\begin{specialsound}[$m$-consumed Type Soundness]
    \label{thm:c-ana-sound}
    If $\gamma, (g, \trace) \vDash_{\mathit{mc}} \Gamma, \graph$ and $\Gamma, \graph \vdash_{\mathit{mc}} e : \type, \graph'$ and $\overline{\psem{e}}_\gamma (g,\trace), w = v, (g', \trace'), w'$, then $v, (g', \trace') \vDash_{\mathit{mc}} t, \graph'$
\end{specialsound}

\begin{proof}
    By structural induction on derivations of $\vdash_{\mathit{mc}}$.
\end{proof}

\newcommand{\append}{\oplus}

Proving the soundness of the $m$-consumed judgment producing the \texttt{bounded} type requires strengthening this theorem to work with partial traces, meaning the abstract graph applies only to the tail end of the trace rather than the whole trace. Using the notation $\trace_1 \append \trace_2$ to mean the trace $\trace_1$ appended with $\trace_2$, we formalize this as follows:

\begin{lemma}{$m$-consumed Soundness on Partial Traces}
    \label{lem:mc-sound-partial}
    If $\gamma, (g, \trace_2) \vDash_{\mathit{mc}} \Gamma, \graph$ and $\Gamma, \graph \vdash_{\mathit{mc}} e : \type, \graph'$ and $\overline{\psem{e}}_\gamma (g,\trace_1 \append \trace_2), w = v, (g', \trace'), w'$, then $\trace' = \trace_1 \append \trace_2'$ and $v, (g', \trace_2') \vDash_{\mathit{mc}} t, \graph'$
\end{lemma}

\begin{proof}
    By structural induction on derivations of $\vdash_{\mathit{mc}}$. The individual steps are the same as the previous theorem, except that they also use the associativity of $\oplus$.
\end{proof}

An unseparated-path type judgment is sound if it abstracts the unseparated path property of the semantics according to the entailment relations.

\begin{specialsound}[Unseparated Path Type Soundness]
    \label{thm:u-ana-sound}
    If $\gamma, (g, \trace) \vDash_{\mathit{up}} \Gamma, \graph$ and $\Gamma, \graph \vdash_{\mathit{up}} e : \type, \graph'$ and $\overline{\psem{e}}_\gamma (g,\trace), w = v, (g', \trace'), w'$, then $v, (g', \trace') \vDash_{\mathit{up}} t, \graph'$
\end{specialsound}

We also strengthen this theorem to aid in proving the soundness of the \texttt{bounded} judgment.

\begin{lemma}{Unseparated Paths Soundness on Partial Traces}
    \label{lem:up-sound-partial}
    If $\gamma, (g, \trace_1 \append \trace_3) \vDash_{\mathit{up}, \mathit{frv}(\trace_1 \append \trace_3)} \Gamma, \graph$ and $\Gamma, \graph \vdash_{\mathit{up}} e : \type, \graph'$ and $\overline{\psem{e}}_\gamma (g,\trace_1 \append \trace_2 \append \trace_3), w = v, (g', \trace'), w'$, then $\trace' = \trace_1 \append \trace_2 \append \trace_3'$ and $v, (g', \trace_1 \append \trace_3') \vDash_{\mathit{up}, \mathit{frv}(\trace_1 \append \trace_3')} t, \graph'$
\end{lemma}

\begin{proof}
    By structural induction on derivations of $\vdash_{\mathit{up}}$. The individual steps are the same as the previous theorem, except that they also use the associativity of $\oplus$.
\end{proof}

\begin{theorem}[Analysis Soundness]
    If $\gamma \vDash_\alpha \Gamma$ and $\Gamma \vdash_\alpha m : \bounded$, then $\overline{\sem{m}}_\gamma \vDash_\alpha \bounded$.
\end{theorem}

\begin{proof}
    We first show that any execution of a stream function $m$ satisfying $\Gamma \vdash_\mathit{mc} m\ \bounded$ satisfies the high-level $m$-consumed semantic property.
    We then show that any execution of a stream function $m$ satisfying $\Gamma \vdash_\mathit{up} m\ \bounded$ satisfies the high-level unseparated paths semantic property.
    Then, by \Cref{thm:hl-ll-equiv} and \Cref{lem:executions}, if $m$ satisfies both these properties, then calling \texttt{infer} on $m$ must be bounded.

    \paragraph{$m$-consumed} Here, we show that any execution of a stream function $m$ satisfying $\Gamma \vdash_\mathit{mc} m\ \bounded$ satisfies the $m$-consumed semantic property. 
        We proceed by induction on the steps of the execution. 
        Note that at each step the program only adds to the trace and thus only adds new introduced variables we ust reason about. 
        We show that the variables introdued at each time step will all be $m$-consumed.
        
        We show this using the definition of $\vdash_\mathit{mc}$. 
        Let $(g_i, \trace_i)$ be the $i$th step of the execution. 
        By \Cref{lem:mc-sound-partial}, $\mathcal{G}$ captures all variables introduced at time $i$.
        Also by \Cref{lem:mc-sound-partial}, $\mathcal{G}'$ captures the variables that are guaranteed to be consumed between $i$ and $i+n$.
        Thus, any variable introduced at time step $i$ must be consumed within $n$ steps (where $n$ is a static bound).
        If it is consumed, the variable will be $m$-consumed at all future time steps where $m$ is at most $n$ times a constant bound based on the number of sample statements in the stream function.

    \paragraph{Unseparated Paths} We proceed by contradiction.
        Assume that $s_i, (g_i, \trace_i)_{i \in \mathbb{N}}$ is an execution that violates the unseparated paths semantic property.
        At some time step $j$, the execution must a) add a variable to the delayed sampling graph in such a way that it increases the unseparated path starting from some variable in the graph, and b) store the variable starting the increased path in $s_{j+1}$.
        Otherwise, the execution would easily satisfy the property.
        According to \Cref{lem:up-sound-partial}, we must have that after each iteration the abstract graph also has a variable with starting an increased path and that some reference $r^*$ contained in the type $\type'$ references this variable.
        Letting $\hat{R}$ be the set of all possible references $r^*$, by the pidgeonhole principle, after $k \ge \size(t) = \size(t') \ge |\hat{R}|$ iterations,\footnote{the equality of $\size(t)$ and $\size(t')$ is enforced by the type rules in \Cref{fig:muF-types-program} of \Cref{sec:muF-types}} the longest path in the abstract graph starting from a variable referenced by an element of $\hat{R}$ must have increased by at least $1$.
        Similarly, after $n \ge \textrm{path}(t, \graph)$ instances of this pattern, the longest path in the abstract graph starting from a variable referenced in $\hat{R}$ must have increased by at least $\textrm{path}(\graph, t)$ and thus be the longest such graph starting from a state variable.
        This contradicts the termination condition that  $\textrm{path}(t, \graph) = \textrm{path}(t'', \graph')$.
\end{proof}


\section{Benchmarks}
\label{sec:benchmarks}

Each of these benchmarks are followed by a \zl{main} stream that serves as the entry point of the program:
\begin{lstlisting}
val main = stream {
  init = infer f;
  step (f, args) = unfold (f, args)
}
\end{lstlisting}

\subsection{Kalman}
\label{ex:kalman}
\begin{lstlisting}
val f = stream {
  init = 0.;
  step (pre_x, obs) =
    let x = sample (gaussian (pre_x, 1.0)) in
    let () = observe (gaussian (x, 1.0), obs) in
    (x, x)
}
\end{lstlisting}

\subsection{Kalman Hold-First}
\label{ex:kalman-first}
\begin{lstlisting}
val kalman = stream {
  init = (true, 0., 0.);
  step ((first, i, pre_x), obs) =
    let (i, pre_x) =
      if first then (let i = sample (gaussian(0., 1.)) in (i, i))
      else (i, pre_x) in
    let x = sample (gaussian (pre_x, 1.)) in
    let () = observe (gaussian (x, 1.), obs) in
    (x, (false, i, x))
}
\end{lstlisting}

\subsection{Gaussian Random Walk}
\label{ex:kalman-generative}
\begin{lstlisting}
val f = stream {
  init = (true, 0.);
  step ((first, x), ()) =
    let x = if first then sample (gaussian (0., 1.)) else sample (gaussian (x, 1.)) in
    (x, (false, x))
}
\end{lstlisting}

\subsection{Coin}
\label{ex:coin}
\begin{lstlisting}
val f = stream {
  init = (true, 0.);
  step ((first, xt), yobs) =
    let xt = if first then sample (beta (1., 1.)) else xt in
    let () = observe (bernoulli (xt), yobs) in
    (xt, (false, xt))
}
\end{lstlisting}

\subsection{Outlier}
\label{ex:outlier}
\begin{lstlisting}
val f = stream {
  init = (true, 0., 0.);
  step ((first, xt, outlier_prob), yobs) =
    let (xt, outlier_prob) =
      if first then
        (sample (gaussian (0., 100.)), sample (beta (100., 1000.)))
      else (sample (gaussian (xt, 1.)), outlier_prob) in
    let is_outlier = sample (bernoulli (outlier_prob)) in
    let () =
      if is_outlier then (observe (gaussian (0., 100.), yobs))
      else (observe (gaussian (xt, 1.), yobs)) in
    (xt, (false, xt, outlier_prob))
}
\end{lstlisting}

\subsection{MTT}
\label{ex:mtt}
\begin{lstlisting}
val f = stream {
  init = (true, List.nil);
  step ((first, t), (inp, cmd)) =
    let last_t = t in
    let t_survived = 
      List.filter (fun (_, _) -> eval (sample (bernoulli (0.5))), last_t) in
    let n_new = sample (poisson (1.0)) in
    let t_new = List.init (n_new, fun _ -> (0, sample (bernoulli (0.5)))) in
    let t_tot = List.append (t_survived, t_new) in
    let t = List.map (fun (tr_num, tr) -> (tr_num, sample (bernoulli (tr))), t_tot) in
    let obs = List.map (fun (_, tr) -> bernoulli (tr), t) in
    let n_clutter = sub (List.length (inp), List.length (obs)) in
    let () = observe (poisson (0.5), n_clutter) in
    let clutter = List.init (n_clutter, fun _ -> bernoulli (tr)) in
    let obs_shuffled = sample (shuffle (List.append (obs, clutter))) in
    let () =  
      if (not (lt (n_clutter, 0))) then
        List.iter2 (fun (var, value) -> 
          observe (gaussian (0.5, var), value), obs_shuffled, inp) 
      else () in
    (t, (false, t))
}
\end{lstlisting}

\subsection{SLAM}
\label{ex:slam}
\begin{lstlisting}
val f = stream {
  init = (true, 0., Array.empty);
  step ((first, x, map), (obs, cmd)) =
    let map = 
      if first then Array.init (100, fun _ -> sample (bernoulli (0.5))) else map in
    let wheel_slip = sample (bernoulli (0.5)) in
    let x = if first then 0. else if wheel_slip then x else plus (x, cmd) in
    let o = Array.get (map, x) in
    let _ = observe (bernoulli (ite (o, 0.9, 0.1)), obs) in
    ((x, map), (false, x, map))
}
\end{lstlisting}

\section{Precision Limitations}
\label{sec:precision-app}

The precision of the analysis is limited by path and complex sensitivity, two common challenges for static analysis. The analysis can be overly conservative when facing conditional branches, for example in the following snippet:
\begin{lstlisting}
let x = sample bernoulli(0.5) in
let y = sample gaussian (0., 1.) in
let () = if x then observe (gaussian (y, 1.), 1.) else () in
let () = if x then () else observe (gaussian (y, 1.), -1.) in
y
\end{lstlisting}
According to the analysis, $\zl{y}$ is not consumed because each branch is separately and conservatively judged to not consume $\zl{y}$, even though there is no path where $\zl{y}$ is unobserved. A more sophisticated analysis that reasons about actual values, not just affected variables, would be more precise here.

Similarly, the analysis can be imprecise in the presence of complex data such as tuples. Consider the following snippet:
\begin{lstlisting}
let x = sample (gaussian(0., 1.)) in
let y = sample (gaussian(0., 1.)) in
let (a, b) =
  if (sample (bernoulli(0.5))) then
    (gaussian (x, 1.), gaussian (y, 1.))
  else (gaussian (y, 1.), gaussian (x, 1.)) in
let () = observe (a, 1.) in
let () = observe (b, 2.) in
(x, y)
\end{lstlisting}
Like the previous example, $\zl{x},\zl{y}$ are not considered consumed even though there is no path that does not observe both. The analysis can determine that both $\zl{a}$ and $\zl{b}$ \emph{may} reference $\zl{x}$ and $\zl{y}$ but neither alone \emph{must} do so. Knowledge about $\zl{a}$ and $\zl{b}$ taken as a pair is lost when they are stored into the tuple. In this case, some kind of alias or shape analysis might recover the relationship between the fields of a tuple.

Without executing for multiple iterations, the $m$-consumed analysis would be occasionally too conservative due to requiring that all variables be used before the end of the current iteration of the step function. Consider:
\begin{lstlisting}
stream {
  init = 0.;
  step (x_prev, obs) =
    let _ = observe (gaussian (x_prev, 1.), obs) in
    let x = sample (gaussian (x_prev, 1.)) in
    (x, x)
}
\end{lstlisting}
In this example, every sample is eventually consumed but only on the subsequent iteration of the step function. If the $m$-consumed analysis only considered one iteration, it would reject this example. Allow introduced variables to be consumed over multiple iterations as we do allows this example to pass the analysis.

Most examples do not require a significant number of iterations for the unseparated paths analysis to converge. However, the analysis may fail to detect convergence in programs with many variables if the iteration bound parameter is too low, as in the following program which requires four iterations:
\begin{lstlisting}
stream {
  init = (0., 0., 0., 0.);
  step ((x_p, x_pp, x_ppp, x_pppp), obs) =
    let x = sample (gaussian (x_p, 1.)) in
    let _ = observe (gaussian (x, 1.), 1.0) in
    (x_pppp, (x, x_p, x_pp, x_ppp))
}
\end{lstlisting}
In this program, the longest unseparated path increases over four iterations, after which variables start being dropped from the state and the maximum length converges.
We suggest that the parameter should be set to be comfortably larger than the number of variables or statements in the program to avoid this issue. Since each iteration is fast to run, it should not cause performance degradation.

Finally, the analysis could incorporate higher-order functions, though they would be hard to analyze statically, and the storage of chains of closures built over many iterations could itself violate a bound on memory usage.

\end{document}